\documentclass[10pt, final, journal, twocolumn, letterpaper]{IEEEtran} 
\usepackage{tikz}
\usepackage{pgfplots}
\usepackage{bm}
\usepackage{mathtools}
\usepackage{graphicx}
\usepackage{relsize}
\usepackage{cite}
\usepackage{amsmath}
\usepackage{amsfonts}
\usepackage{color}
\usepackage{xcolor}
\usepackage{psfrag}
\usepackage{amssymb}
\usepackage{dblfloatfix}
\usepackage[font=small]{caption}
\usepackage{tabularx}
\usepackage{makecell}
\usepackage{lipsum}
\usepackage{float}
\usepackage{subfiles}
\usepackage{epstopdf}
\usepackage[inline]{enumitem}
\usepackage{verbatim}
\usepackage{subfig}
\usepackage{scalerel}

\newcommand{\RomanNumeralCaps}[1]{\MakeUppercase{\romannumeral #1}}
\def \fatR {{ \mathbb R}}

\def\rr{{Rx}}
\def\s{{Tx}}

\def \A {{\bm A}}
\def \B {{\bm B}}

\def \D {{\bm D}}
\def \F {{\bm F}}
\def \H {{\bm H}}
\def \K {{\bm K}}
\def \J {{\bm J}}
\def \I {{\bm I}}
\def \R {{\bm R}}
\def \Q {{\bm Q}}
\def \P {{\bm P}}

\providecommand{\norm}[1]{\lVert#1\rVert}
\newtheorem{problem}{\noindent \textbf{Problem}}

\IEEEoverridecommandlockouts

\begin{document}
\allowdisplaybreaks
\frenchspacing

\title{Massive MIMO-based Localization and \\ Mapping Exploiting Phase Information of \\ Multipath Components \vspace*{-1mm}}

\author{\normalsize Xuhong Li,~\IEEEmembership{\normalsize Student Member,~IEEE}, Erik Leitinger,~\IEEEmembership{\normalsize Member,~IEEE},\\ Magnus Oskarsson,~\IEEEmembership{\normalsize Member,~IEEE}, Kalle {\AA}str\"{o}m,~\IEEEmembership{\normalsize Senior Member,~IEEE},
\\Fredrik Tufvesson,~\IEEEmembership{\normalsize Fellow,~IEEE\vspace*{-3mm}}
\thanks{Manuscript received October 15, 2018; revised March 18, 2019; accepted May 28, 2019.  This work was supported in part by the Swedish Research Council (VR), in part by ELLIIT an Excellence Center, in part by the e-Science Collaboration (eSSENCE), and in part by the Austrian Science Fund (FWF) under the grant J 4027. The associate editor coordinating the review of this paper and approving it for publication was D.\,Matolak. (\textit{Corresponding author: Xuhong Li.})\vspace{-3mm}\newline

X.\ Li, and F.\ Tufvesson are with the Department of Electrical and Information Technology, Lund University, SE-221 00 Lund, Sweden (e-mail: \{xuhong.li, fredrik.tufvesson\}@eit.lth.se).\vspace{-3mm}\newline 

E.\ Leitinger is with the Signal Processing and Speech Communication Laboratory, Graz University of Technology, A-8010 Graz, Austria (e-mail: erik.leitinger@tugraz.at).\vspace{-3mm}\newline

M.\ Oskarsson and K.\ {\AA}str\"{o}m are with the Centre for Mathematical Sciences, Lund University, Lund, Sweden (e-mail: \{magnuso, kalle\}@maths.lth.se).
}\vspace{0mm}}

\maketitle

\begin{abstract}
In this paper, we present a robust multipath-based localization and mapping framework that exploits the phases of specular multipath components (MPCs) using a massive multiple-input multiple-output (MIMO) array at the base station. Utilizing the phase information related to the propagation distances of the MPCs enables the possibility of localization with extraordinary accuracy even with limited bandwidth. The specular MPC parameters along with the parameters of the noise and the dense multipath component (DMC) are tracked using an extended Kalman filter (EKF), which enables to preserve the distance-related phase changes of the MPC complex amplitudes. The DMC comprises all non-resolvable MPCs, which occur due to finite measurement aperture. The estimation of the DMC parameters enhances the estimation quality of the specular MPCs and therefore also the quality of localization and mapping. 
The estimated MPC propagation distances are subsequently used as input to a distance-based localization and mapping algorithm. This algorithm does not need prior knowledge about the surrounding environment and base station position. The performance is demonstrated with real radio-channel measurements using an antenna array with 128 ports at the base station side and a standard cellular signal bandwidth of 40\,MHz. The results show that high accuracy localization is possible even with such a low bandwidth.

\end{abstract}



\IEEEpeerreviewmaketitle

\section{Introduction}
\label{s1}

High precision localization is a key enabler for future location-aware applications expected in future 5G communication networks \cite{DiTaranto2014SPM}. Therefore, localization techniques that can offer the necessary accuracy in complex environments, e.g., dense urban environments or indoors, are strongly needed. Massive multiple-input-multiple-output (MIMO) transmission schemes \cite{RusekSPM2013}, \cite{NilGarcia_Trans2017} are one possibility to counteract localization degradation due to harsh multipath propagation in dense urban environments and indoors, even though only small signal bandwidth is used.

\subsection{State of the Art}

Achieving the required level of accuracy robustly is still elusive in environments that are characterized by harsh multipath channel conditions. Therefore, most existing localization approaches supporting multipath channels either use sensing technologies that mitigate multipath effects \cite{StefanoMarano_Journal2010, WymeerschTC2012, YWang_Globecom2017} or fuse information from multiple information sources \cite{YuanShen_Trans2010, ShenJSAC2012, Buehrer_Proceeding2018, MoeWin_Proceeding2018}. Fingerprint-based approaches actually exploit the diversity of multipath channels by matching position-labeled channel measurements with the acquired measurements at the positions of interest \cite{RSSI_ref2, SteinerTSP2011}. 
Similarly, this can be achieved by using machine learning methods with the additional capability of interpolation between the position-labeled channel training measurements \cite{SavicVTC2015_Fingerprinting, Joao_fingerprint_2017}. However, site-specific training phases require a lot of accurate position-labeled channel measurements and may lead to performance degradation in dynamic environments.

Multipath-assisted localization algorithms \cite{LeitingerJSAC2015, Klaus_Trans2016, ShahmansooriTWC2018, MmWave_bounds_Henk} exploit position-related information contained in the specular multipath propagation components (MPCs) that can be associated to the local geometry, which actually turns the multipath effect into an advantage. MPCs due to specular reflections at flat surfaces are modeled by virtual anchors (VAs), which are mirror images of the physical anchors (PAs), i.e., base station \cite{Borish1984}. By associating the estimated MPC parameters to VAs, these VAs can be used as additional PAs for location estimation. In recent years, many works that use wideband/ultra-wideband (UWB) signals have shown the potential of multipath-based positioning \cite{LeitingerICC2014}, tracking \cite{LeitingerGNSS2016, MeissnerWCL2014} and simultaneous localization and mapping (SLAM) \cite{GentnerTWC2016, KuangICC13, LeitingerICC2017, LeitingerArxiv2018} with accuracy on a centimeter level. The works \cite{FroehleICC2013_BPCoMINT, KulmerTWC2018_CoopMINT, Naseri_TSP2018} use cooperation amongst agents to enhance multipath-assisted localization performance in infrastructure-limited scenarios. 

However, all these multipath-assisted algorithms have in common that they require accurate extraction of location-related parameters of MPCs (i.e., distances/delays and angles). The estimation quality of MPC parameters in turn determines the localization performance, while, good resolvability between MPCs is a prerequisite for accurate estimation. However, using only limited bandwidth systems leads to low resolvability of MPCs in the delay domain, especially in dense multipath environments. Utilizing large-scale antenna arrays extends signal processing alternatives from the time-frequency domain to also include the spatial domain, and therefore helps to resolve closely spaced MPCs by exploiting the spatially sparse structure of the multipath channel \cite{WitrisalWCL2016}. In \cite{ShahmansooriTWC2018, MmWave_bounds_Henk}, the theoretically achievable localization performance given as the Cram{\'e}r-Rao lower bound (CRLB) on the position and orientation estimation error for millimeter-wave massive MIMO systems is presented. The results in there show the large localization performance improvement when position-related information of MPCs is estimated with a massive MIMO system. Considering that cellular systems are typically operating at a few GHz with a bandwidth of 20-40\,MHz, the corresponding resolution of one time sample is only 7.5-15\,m. However, since the phase of MPCs is connected to the carrier frequency and this lies in the GHz region for typical radio systems, centimeter accuracy can be achieved if the phase is properly exploited as for example in global navigation satellite systems \cite{Phase_ref1} or terrestrial radio systems \cite{ZhuICCW2015}. If the spatial sampling rate of the radio channel is sufficiently high, i.e., recording a few snapshots within one wavelength movement, it is possible to track the distance change on centimeter level by measuring the phase shift between measurements at two consecutive time instances. 

\subsection{Contributions and Organization of the Paper}

In this work, a multipath-assisted localization and mapping framework is presented that exploits the phase information of individual MPCs by using a massive single-input multiple-output (SIMO)\footnote{We consider a simpler scenario here, where the mobile agent is equipped with a single omnidirectional antenna. However, the framework can be easily extended to MIMO setup.} radio system. As shown in Fig.~\ref{Principle}, the framework is composed of two consecutive steps: (i) Using an extended Kalman filter (EKF) \cite{SalmiTSP2009}, which is initialized with the iterative maximum-likelihood estimation algorithm (RIMAX) \cite{RichterPhD2005}, the dispersion parameters, i.e., the delays/distances and angle-of-arrivals (AoAs) of the specular MPCs, the noise and dense multipath component (DMC) parameters are estimated from channel measurements; (ii) The MPC distances estimates are subsequently used as input in the localization step\footnote{In the SIMO setup, the angular information is available at PA side. The angular information is exploited by the channel estimator for better resolvability of individual MPCs, however, in this work the AoAs are not used for localization.}. A distance-based algorithm is applied that simultaneously estimates the mobile agent positions and VA/PA positions\footnote{Since these estimates are only relative w.r.t. a global coordinate system, we register them to the coordinate system of the measured geometric groundtruth of the mobile agents' movement trajectory.}. Both synthetic and real channel measurements are used when demonstrating the performance of MPC parameter estimation, followed by an in-depth statistical analysis of MPC parameters in terms of lifetime, signal-to-interference-plus-noise ratio (SINR)\footnote{The SINR can be interpreted as a reliability measure of estimated MPC parameters and is directly tied to the CRLB of multipath-based localization \cite{LeitingerJSAC2015}.}, etc. The performance of the localization algorithm is evaluated with the same real channel measurements.    

The main contributions are summarized as: 
\begin{itemize}
\item We present a novel MIMO channel estimation and tracking algorithm that tightly couples the tracked MPC distances to the phase change of the MPC complex amplitudes from one measurement snapshot to the next. With this, it is possible to estimate the MPC distances far beyond the signal bandwidth dependent accuracy.
\item We analyze the dynamic behaviour and statistical distributions of the estimated MPC parameters and connect them to the localization potential. 
\item We use the estimated MPC distances from real channel measurements to show that radio-based localization in harsh multipath environments is possible even using only low signal bandwidth by exploiting a massive MIMO system. 
\end{itemize}
Parts of this paper were published in \cite{Xuhong_PIMRC_2017}, where the feasibility of the phase-based localization using standard cellular bandwidths was demonstrated. This paper presents more insights into the framework as well as more in-depth analysis of the channel estimation results with both synthetic and real measurements.

The rest of the paper is structured as follows: Section~\ref{s2} introduces the radio signal model and the multipath-based localization and mapping problem. Section~\ref{s3} and \ref{s4} present EKF-based channel estimation and tracking algorithm and distance-based localization and mapping algorithm. The numerical results and analysis are reported in Section~\ref{s5}. Finally, Section~\ref{s6} concludes the paper. 

\textit{Mathematical notations}: Boldface upper case letters represent matrices. Boldface lower case letters denote column vectors. Superscripts $^{\text{T}}$, $^\ast$ and $^{\text{H}}$ denote matrix transpose, complex conjugation and Hermitian transpose, respectively. The Kronecker product and Khatri-Rao product operators are denoted with $ \otimes $ and $ \lozenge $, respectively. $ \norm{\cdot} $ is the Euclidean norm. $ \vert\cdot\vert $ represents the absolute value. $ \text{card}(\cdot) $ denotes the cardinality of a set. $ \hat{\textbf{A}} $ denotes an estimate of $ \textbf{A} $. $ \I_{[\,\bm{\cdot}\,]} $ represents identity matrix with dimension denoted in the subscript $[\,\bm{\cdot}\,]$. $ \text{diag}(\bm{a}) $ denotes a diagonal matrix with the vector $ \bm{a} $ being the diagonal entries. The operation $ \text{toep}(\bm{a}, \bm{a}^{\text{H}}) $ constructs a Hermitian Toeplitz matrix with vectors $ \bm{a} $ and $ \bm{a}^{\text{H}} $ being the first column and the first row.


\section{Problem Formulation}
\label{s2}
\begin{figure}[t!]
	\centering
	\includegraphics[width=0.45\textwidth]{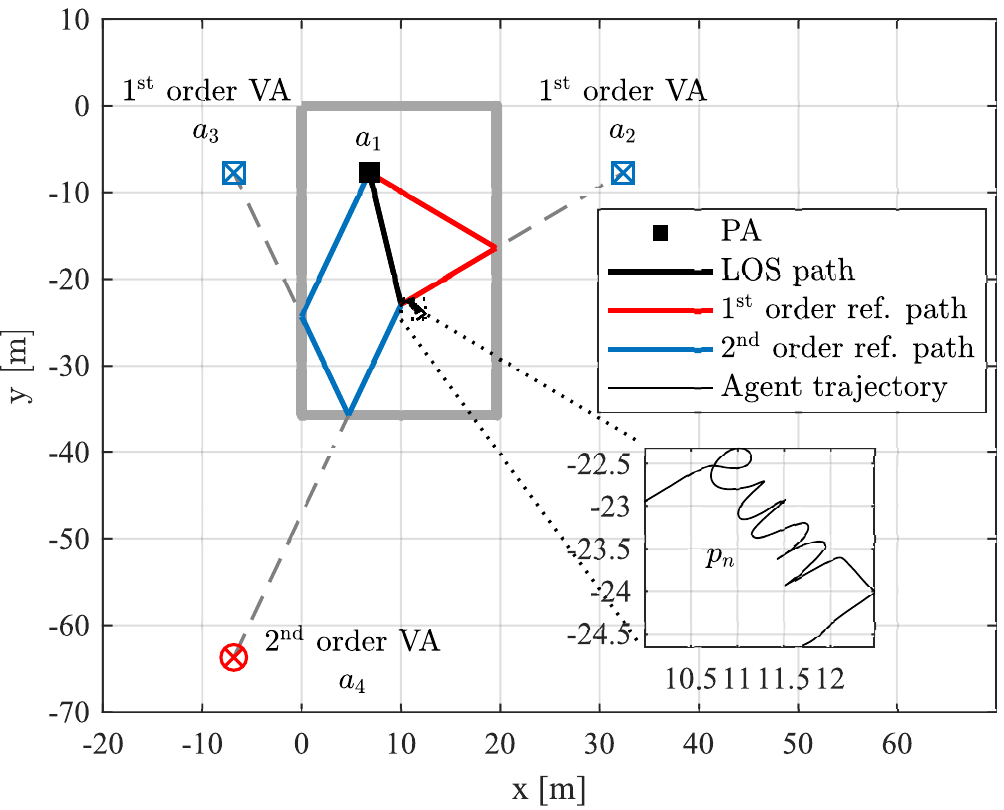}
	\caption{Floor plan of the sports hall in Medicon village, Lund, Sweden. The bold grey line represents the surrounding walls. Besides, three examples of the $ 1^{\text{st}} $ order and $ 2^{\text{nd}} $ order geometrically expected VAs, as well as the corresponding reflection paths from the mobile agent to the physical anchor (PA) are given. The groundtruth trajectory of the mobile agent is given by the letters ``\textbf{Lund}'' in a 2\,$ \text{m}^2 $ area, as shown in the zoom-in sub-plot.}
	\label{Theory}
	\vspace{0pt}
\end{figure}

Multipath-based localization utilizes geometrical information contained in specular MPCs---delays/distances, angles of departure (AoDs) and angles of arrival (AoAs)---estimated from received radio signals \cite{LeitingerJSAC2015}. Each estimated specular MPC, which originates from a reflection on planar surfaces, can be either associated with a PA or with one of the VAs, which represent the mirrored positions of the PA w.r.t. the planar surfaces. These VAs can be used as additional PAs for localization. From now on a PA or VAs are collectively referred to as features. Fig.~\ref{Theory} shows the floor plan of the indoor environment in which the measurement campaign was performed, together with the positions of the PA and of three exemplary VAs with their corresponding reflection paths. We consider the case that the mobile agent acts as a transmitter with unknown time-varying positions $ \bm{p}_n \in \mathbb{R}^{3 \times 1} $, $ n = 2, \dots, N $. The feature positions are denoted with $ \bm{a}_{m} \in \mathbb{R}^{3 \times 1} $, $ m \in \mathcal{M} = \lbrace 1,\dots, M\rbrace $, where the PA represents a receiver at a static but unknown position $ \bm{a}_{1} $, and the positions\footnote{The coordinate of the position is given in the 3D Cartesian coordinate system. For a feature position, $ \bm{a}_{m} = [x_{m}\quad y_{m}\quad z_{m}]^{\text{T}} $, and for the mobile agent position $ \bm{p}_{n} = [x_{n}\quad y_{n}\quad z_{n}]^{\text{T}} $. } of the geometrically expected VAs are denoted as $ \bm{a}_{m} $, $ m = 2,\dots, M $. The feature positions are fixed over time since the PA is static. A specular MPC is consistently associated with a feature for the duration that this feature is visible at the agent position. Those visible features at each agent position $ \bm{p}_n $ are called expected features, with the positions given as $ \bm{a}_{l} \in \mathbb{R}^{3 \times 1} $, $ l \in \mathcal{L}_{n} $, and $ \mathcal{L}_{n} $ is a subset of $ \mathcal{M} $, i.e., $ \mathcal{L}_{n}\subseteq \mathcal{M} $. The number of expected features $ L_{n} = \text{card}(\mathcal{L}_{n})$ is unknown and time-varying, and it depends on the visibility at each agent position. Besides, the floor plan of the surrounding environment is assumed as unknown, which means feature positions $ \bm{a}_{l} $ are unknown.

\begin{figure}[t!]
	\centering
	\includegraphics[width=.45\textwidth]{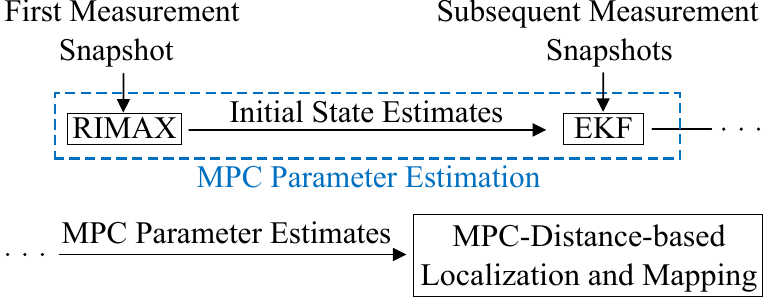}
	\caption{Block diagram of the proposed multipath-based localization and mapping framework.}
	\label{Principle}
	\vspace{0pt}
\end{figure}

Fig.~\ref{Principle} shows the block diagram of the proposed multipath-based localization and mapping framework. First, the MPC parameters are estimated using an EKF-based channel estimator and tracking algorithm. Considering that an accurate initial state estimate is a prerequisite for the fast convergence and accurate tracking in the EKF, and the initialization step should avoid bringing too many artifacts into the initial state vector, the RIMAX algorithm is applied at time $ n = 1 $ for the initial estimates of MPC parameters and noise covariance \cite{RichterPhD2005}. The estimated MPC distances are subsequently used in the localization and mapping algorithm. 

\subsection{Radio Signal Model} 
The baseband signal $ \bm{y}_n $ in frequency domain received by the PA at time $n$ is modeled as
\begin{align}
\bm{y}_n = \bm{s}_n + \bm{w}_{\text{dmc},n} + \bm{w}_n \quad \in \mathbb{C}^{ N_\text{f} N_{\text{Tx}} N_{\text{Rx}}\times 1},
\label{eq:Received_model} 
\end{align}
where the first term comprises specular MPCs and the second and third terms represent DMC and additive white Gaussian noise, respectively. We assume time-synchronization between the mobile agent and the PA, and time synchronization between VAs is automatically achieved as they arise due to reflections. The values $ N_\text{f} $, $ N_{\text{Tx}} $ and $ N_{\text{Rx}} $ refer to the number of frequency sample points, transmit and receive antenna elements, respectively. Since $ N_{\text{Tx}} = 1$ for a SIMO setup, we ignore $ N_{\text{Tx}} $ in the dimension notations of matrices/vectors from now on. 

\subsubsection{Specular MPCs} 
$ \bm{s}_n = [ \bm{s}_{f_1,n}^{\text{T}}\cdots\bm{s}_{f_{N_{\text{f}}},n}^{\text{T}}]^{\text{T}} $ is obtained by sampling the continuous response $ \bm{s}_n(f) $ in the frequency domain at time $ n $, where $ f_i $ with $i = 1,\dots, N_{\text{f}} $ is the frequency samples in the domain $ \lbrace -\frac{N_{\text{f}}-1}{2N_{\text{f}}}B_{\text{w}},\dots,\frac{N_{\text{f}}-1}{2N_{\text{f}}}B_{\text{w}} \rbrace $ and $ B_{\text{w}} $ is the signal bandwidth. $ \bm{s}_n(f) = \sum_{l \in \mathcal{L}_{n}}\bm{s}_{l,n}(f) $ comprises $ L_n $ specular MPCs. The contribution of each MPC is given by $ \bm{s}_{l,n}(f) = s_{\text{sig},n}(f)\bm{h}_{l,n}(f) $, where $ s_{\text{sig},n}(f)$ is the transmitted baseband signal response, and $ \bm{h}_{l,n}(f) \in \mathbb{C}^{N_{\text{Rx}}\times 1}$ is a frequency domain representation of the MPC's channel impulse response, given as \cite{Molisch:2011:WC:1984860} 
\begin{align}
\bm{h}_{l,n}(f)=\B_{\text{Rx}}(\varphi_{l,n},\theta_{l,n}) \bm{\Gamma}_{l,n}\B_{\text{Tx}}^{\text{T}}\text{e}^{-j2\pi (f+f_{\text{c}})\tau_{l,n}},
\label{eq:double_direc_model} 
\end{align}
where $ f_c $ is the carrier frequency, $ \tau_{l,n} $ represents the propagation delays of the $l$th MPC. The matrices $ \B_{\text{Tx}}\in \mathbb{C}^{1\times 2}$ and $ \B_{\text{Rx}}(\cdot,\cdot) \in \mathbb{C}^{N_{\text{Rx}}\times 2} $ describe the far-field complex antenna responses of the omnidirectional antenna at the transmit side, and of the antenna array at the receive side w.r.t. the AoAs in elevation and azimuth domain, respectively. The delay of the specular MPC indexed by $ l $ is proportional to the distance between the agent and the PA or between the agent and the VAs. That is $ \tau_{l,n} = \norm{\bm{p}_n-\bm{a}_{l}}/c = d_{l,n}/c $, where $d_{l,n}$ is the propagation distance and $ c $ is the speed of light. We note that similar geometrical expressions can be extended to the azimuth and elevation AoAs $ (\varphi_{l,n},\theta_{l,n}) $, respectively. The parameters of each specular MPC are comprised in the vector $\bm{\mu}_{l,n} = [d_{l,n} \quad \varphi_{l,n} \quad \theta_{l,n}] \in \mathbb{R}^{3\times 1}$. The complex path weight matrix $ \bm{\Gamma}_{l,n} \in \mathbb{C}^{2\times 2} $ accounts for the frequency independent attenuation and phase change, given as 
\begin{equation}
\bm{\Gamma}_{l,n} =
\begin{bmatrix}
\gamma_{\text{HH},l,n} & \gamma_{\text{VH},l,n}\\
\gamma_{\text{HV},l,n} & \gamma_{\text{VV},l,n}
\end{bmatrix}.
\end{equation}
The individual polarimetric complex path weights of the matrix are given by $ \gamma_{\text{p},l,n} =\alpha_{\text{p},l,n} e^{j\phi_{\text{p},l,n}} $, where $ \alpha_{\text{p},l,n} $ and $ \phi_{\text{p},l,n} $ represent the magnitude and phase, respectively. The subscript $ \text{p}\,\in\,\lbrace\text{{\footnotesize HH}}, \text{{\footnotesize HV}}, \text{{\footnotesize VH}}, \text{{\footnotesize VV}}\rbrace $ indicates the four polarimetric transmission coefficients (as for example $ \text{{\footnotesize HV}} $ indexes the horizontal-to-vertical transmission coefficient). 
\subsubsection{Noise process}\label{subsec_NoiseVariance} The second term in (\ref{eq:Received_model}), $ \bm{w}_{\text{dmc},n} $ denotes the multiplication of the signal spectrum with the DMC defined by the covariance matrix $ \bm{R}_{\text{dmc},n} \in \mathbb{C}^{ N_\text{f} N_{\text{Rx}}\times N_\text{f} N_{\text{Rx}} } $, and the third term in (\ref{eq:Received_model}), $ \bm{w}_n $ denotes the measurement noise which is assumed circularly symmetric complex Gaussian noise with covariance matrix $\R_{\text{w},n} = \sigma_{\text{w},n}^{2}\I_{N_\text{f} N_{\text{Rx}}} \in \mathbb{C}^{ N_\text{f} N_{\text{Rx}}\times N_\text{f} N_{\text{Rx}} }$. The noise covariance matrix is given by $\R_n = \R_{\text{w},n} + \bm{R}_{\text{dmc},n}$.

The estimation of the noise parameters directly for the noise covariance matrix $\R_n$ is computationally very expensive, especially for the massive MIMO setup. Using the narrowband assumption, a Kronecker decomposition of the noise covariance matrix $\R_n$ can be applied \cite{WeichselbergerPhD2003, RichterPhD2005}. The noise covariance matrix then reduces to
\begin{equation}\label{eq:fullCov}
\R_n = \R_{\text{Rx},n} \otimes \R_{\text{f},n}(\bm{x}_{\text{dmc},n}) + \sigma_{\text{w},n}^2\I_{N_\text{f} N_{\text{Rx}}},
\end{equation}
where $ \R_{\text{f},n} \in \mathbb{C}^{N_\text{f} \times N_\text{f}} $ denotes the covariance matrix of DMC in frequency domain, which has Toeplitz structure and is given as
\begin{equation}
\R_{\text{f},n}(\bm{x}_{\text{dmc},n}) = \text{toep}\left(\bm{\kappa(\bm{x}_{\text{dmc},n})}, \bm{\kappa(\bm{x}_{\text{dmc},n})}^{\text{H}} \right).
\label{eq:TopeRf}
\end{equation}
Here, $ \bm{\kappa(\bm{x}_{\text{dmc},n})} $ is the sampled power delay spectrum (PDS) in frequency domain characterized by $\bm{x}_{\text{dmc},n} = [\alpha_{\text{dmc},n} \quad \beta_{\text{dmc},n}\quad \tau_{\text{on},n}]^{\text{T}}$. The DMC is modeled with an exponentially decaying power profile in the delay domain, where $ \alpha_{\text{dmc},n} $ is the power at the onset delay $\tau_{\text{on},n}$, and $ \beta_{\text{dmc},n} $ is the normalized coherence bandwidth of DMC (detailed parameters can be found in \cite[Section 2.5]{RichterPhD2005}). Furthermore, it is assumed that the DMC is spatially uncorrelated at the receiver side, therefore the covariance matrix in the angular domain $ \R_{\text{Rx},n} = \I_{N_{\text{Rx}}}$.
\subsubsection{Signal Parameter Estimation} 
Given the radio signal observations $ \bm{y} = [\bm{y}_{1}^{\text{T}}\cdots\bm{y}_{N}^{\text{T}}]^{\text{T}} $, the EKF-based parametric channel estimation algorithm, described in Section~\ref{s3}, provides $\widetilde{K}$ continuously estimated MPCs and noise parameters. Those MPCs are indexed by $ \tilde{k} $ with $ \tilde{k} \in \widetilde{\mathcal{K}}= \lbrace 1,\dots,\widetilde{K}\rbrace $, and each of them is consistently associated with an estimated feature position $ \bm{a}_{\tilde{k}} $ for the duration of its lifetime (described in Section~\ref{s_sda}). The estimated MPCs have different lifetimes, which means they are observed during different fractions of the measurement time. At time $ n $, a subset of MPCs indexed by $ k \in \mathcal{K}_{n} $, $ \mathcal{K}_{n} \subseteq \widetilde{\mathcal{K}}$ are estimated, and the estimated parameters of each MPC and the noise parameters are given as 
\begin{align}
\hat{\bm{\mu}}_{k,n} &= [\hat{d}_{k,n} \quad \hat{\varphi}_{k,n} \quad \hat{\theta}_{k,n}]^{\text{T}} \,\in\, \mathbb{R}^{3\times 1}\\
\hat{\bm{x}}_{\text{noise,n}} &= [\hat{\bm{x}}_{\text{dmc},n}^{\text{T}}\quad \hat{\sigma}_{\text{w},n}]^{\text{T}}\,\in\, \mathbb{R}^{4\times 1}.
\end{align}
Ideally, the number of estimated MPCs at time $ n $, i.e., $ K_{n} = \text{card}(\mathcal{K}_{n}) $, should be equal to $L_{n}$. However, during the estimation process, miss detection of specular MPCs and false alarm which leads to clutter components might happen. Hence, $K_{n}$ is time-varying and it can be equal to, or larger/smaller than $L_{n}$. In the next section, the estimated parameters are provided as input to the localization and mapping algorithm.

\subsection{Localization and Mapping Problem}



In this work, we only use the estimated distances $\hat{d}_{k,n}$ of the MPCs within the localization and mapping algorithm. The estimates are corrupted by noise and possible biases, so the measurement model of the localization algorithm is given as 
\begin{equation}
\hat{d}_{k,n} = \norm{\bm{p}_n-\bm{a}_{k}} + \epsilon_{k,n}, \forall (k,n) \in I,   
\label{eq:distance}
\end{equation} 
where $ I $ of all $ (k,n) $ indexing combinations represents the set of all the estimated MPCs. Distance estimates which are considered to be inliers have a known distribution $\epsilon_{k,n} \sim \mathcal{N}(0,\sigma_{\text{inl}}^2)$ for $(k,n) \in I_{\text{inl}}$. Outliers (comprise false alarms and specular MPCs with large errors) represent distance estimates $\hat{d}_{k,n}$ that follow an unknown distribution of $\epsilon_{k,n} $ for $(k,n) \in I_{\text{outl}}$ with typically much larger variance. One useful approach is to minimize the negative log-likelihood. To simplify the problem, we assume that the negative log-likelihood for the outliers is a constant $C$, i.e., each outlier has the same penalty. In this way the problem becomes an optimization problem. 

\noindent \begin{problem} \label{prob_optmisstoa}
({Localization and Mapping}) Given absolute distance estimates $\hat{d}_{k,n} \forall~(k,n)~\in~I$, find the inlier set $I_{\text{inl}}~\subset~I$, the estimated mobile agent positions $\hat{\bm{p}}_n~\forall~n \in \lbrace 1,\dots,N \rbrace $ and the estimated feature positions $\hat{\bm{a}}_{k} \forall~(k,n)~\in~I_{\text{inl}} $ that solves the following optimization problem 
\begin{align}
\min_{I_{\text{inl}},\bm{p}_n,\bm{a}_{k} \atop \forall n, \forall k}\, \sum_{(k,n) \in I_{\text{inl}}} (\hat{d}_{k,n}- \norm{\bm{p}_n-\bm{a}_{k}} )^2 + \sum_{(k,n) \in I_{\text{outl}}} C,  
\label{eq:optim}
\end{align}
\end{problem}
where $I_{\text{outl}} = I \setminus I_{\text{inl}}$. The estimated feature positions $\hat{\bm{a}}_{k} $ are also assumed to be fixed over time since the PA is static. This is a highly non-linear, non-convex optimization problem. The problem changes character if both $\hat{\bm{p}}_n$ and $\hat{\bm{a}}_{k}$ span 3D, or either one of them or both are restricted to a plane or a line as shown in \cite{burgess-kuang-etal-sp-15}. The problem is ill-defined if there is too little data. For planar problems we require $K_n~\geq~3, N~\geq~3$, \cite{stewenius-phd-2005}. For 3D problems more data is needed, typically $K_n~\geq~4, N~\geq~6$ or $K_n~\geq~5, N~\geq~5$, \cite{kuang-burgess-etal-icassp-13}. 

In the following two sections, we introduce the framework as shown in Fig.~\ref{Principle}. At first, the EKF channel estimation and tracking algorithm is presented, followed by the MPC-distances-based localization and mapping algorithm.

\section{EKF-based Channel Estimation Algorithm}
\label{s3}
The MPC parameters are firstly initialized with the RIMAX algorithm, and then an EKF is adopted for continuous channel parameters tracking. It should be noted that instead of estimating the absolute phase of each MPC at each time instance, we track the continuous phase changes between consecutive snapshots. Given a few snapshots being taken within one wavelength movement of the mobile agent, a phase change from 0 to $ 2\pi $ is translated into a distance change $ \Delta d_{l,n} $ from 0 to $ \lambda $. The two parameters $ \bm{\phi}_{\text{p},l,n} $ and $ d_{l,n} $ in (\ref{eq:double_direc_model}) are both phase related, but the estimates $ \hat{\bm{\phi}}_{\text{p},l,n} $ are usually non-continuous in complex propagating environments, which leads to a high risk of phase slip, i.e., a jump of an integer number of phase cycles. Therefore, $ \hat{\bm{\phi}}_{\text{p},l,n} $ of each MPC is locked to the initial
estimate provided by the RIMAX algorithm, the evolution $ \Delta\hat{\bm{\phi}}_{\text{p},l,n} $ however is not involved in the tracking process using the EKF. In detail, we exclude $ \Delta{\bm{\phi}}_{\text{p},l,n} $ from the state space model and the corresponding derivative $ \frac{\partial \bm{s}(\hat{\bm{x}}_{n})}{\partial (\Delta\hat{{\bm{\phi}}}_{n})^\text{T}} $ from the Jacobian matrix (Section~\ref{EKF_mea_update}). In this way, we ensure the unique mapping between the phase shift and the distance change $ \Delta d_{l,n} $. 
 
\subsection{State Space and Measurement Model}
\label{s_SSM}

The state space vector of $L_n$ MPC parameters at time $n$ is given by
\begin{equation}
\bm{x}_n = [\bm{\mu}_n^\text{T} \quad \Delta\bm{\mu}_n^\text{T} \quad \bm{\alpha}_n^\text{T} \quad \bm{\phi}_n^\text{T}]^\text{T} \in \mathbb{R}^{14 L_n\times 1}\, ,
\label{eq:statevector}
\end{equation}
where the geometry-related parameter are stacked into
\begin{align}
\bm{\mu}_n = [\bm{d}_n^{\text{T}} \quad \bm{\varphi}^{\text{T}}_n \quad \bm{\theta}^{\text{T}}_n]^{\text{T}} \in \mathbb{R}^{3 L_n\times 1},
\label{eq:vectorsum}
\end{align}
and the vector $ \Delta\bm{\mu}_n \in \mathbb{R}^{3 L_n\times 1} $ contains the change rates of the MPC parameters contained in $ \bm{\mu}_n $. The magnitudes and phases of the according complex MPC weights are stacked into
\begin{align}
\bm{\alpha}_n &= [\bm{\alpha}_{\text{HH},n}^{\text{T}} \quad \bm{\alpha}_{\text{HV},n}^{\text{T}} \quad \bm{\alpha}_{\text{VH},n}^{\text{T}} \quad \bm{\alpha}_{\text{VV},n}^{\text{T}}]^{\text{T}} \in \mathbb{R}^{4 L_n\times 1},\label{alphavector} \\[0.3em]
\bm{\phi}_n &= [\bm{\phi}_{\text{HH},n}^{\text{T}} \quad \bm{\phi}_{\text{HV},n}^{\text{T}} \quad \bm{\phi}_{\text{VH},n}^{\text{T}} \quad \bm{\phi}_{\text{VV},n}^{\text{T}}]^{\text{T}} \in \mathbb{R}^{4 L_n\times 1}.\label{phivector}
\end{align}
Each sub-vector on the right side of (\ref{eq:vectorsum}), (\ref{alphavector}) and (\ref{phivector}) has the dimension of $ (L_n \times  1) $, as for example, $ \bm{d}_n\, =\,[d_{1,n} \, \cdots \, d_{L_n,n}]^{\text{T}} $, $ \bm{\alpha}_{\text{HH},n} = [\alpha_{\text{HH},1,n}\, \cdots \, \alpha_{\text{HH},L_n,n}]^{\text{T}} $ and $ \bm{\phi}_{\text{HH},n} = [\phi_{\text{HH},1,n} \, \cdots \, \phi_{\text{HH},L_n,n}]^{\text{T}} $.


%

The state transition model defined by a discrete white noise acceleration model \cite[Section 6.3.2]{BarShalom2002EstimationTracking} describes the time evolution of the state vector. With the assumption that the motion and underlying noise process of different MPC parameters are uncorrelated, the discrete-time state transition model is given as
\begin{equation}
\bm{x}_n = \F\bm{x}_{n-1} + \bm{v}_{n},
\label{eq:vector_trans}
\end{equation}
where $ \bm{v}_{n} $ is state noise vector following zero mean normal distribution with the variance matrix $ \Q $. The state transition matrix $ \F^1 \in \mathbb{R}^{14 \times 14} $ of a single MPC is formulated as
\begin{equation} 
\F^1 = 
\begin{bmatrix}
\I_3 & \I_3\Delta T & \bm{0} & \bm{0}\\
 \bm{0} & \I_3 & \bm{0} & \bm{0}\\
\bm{0} & \bm{0} & \I_4 & \bm{0}\\
  \bm{0} & \bm{0} & \bm{0} & \I_4
\end{bmatrix},
\label{eq:transf_matrix}
\end{equation}
where $ \Delta T $ is the channel sampling duration. The variance matrix $ \Q^1 \in \mathbb{R}^{14 \times 14}  $ of a single MPC is defined as
\begin{equation} 
\Q^1 = 
\begin{bmatrix}
 \Q^1_{\bm{\mu}} & \bm{0} & \bm{0}\\
 \bm{0} & \Q^1_{\bm{\alpha}} & \bm{0} \\
\bm{0} & \bm{0} & \Q^1_{\bm{\phi}}
\end{bmatrix}.
\label{eq:Q_matrix}
\end{equation}
The sub-matrix $ \Q^1_{\bm{\mu}} \in \mathbb{R}^{6 \times 6} $ related to the structural vector $ \hat{\bm{\mu}} $ is given as
\begin{equation} 
\Q^1_{\bm{\mu}} = \text{diag}(\bm{q}_{\Delta\bm{\mu}})\otimes
\begin{bmatrix}
 \frac{1}{4}\Delta T^4 & \frac{1}{2}\Delta T^3 \\[0.7em]
 \frac{1}{2}\Delta T^3 & \Delta T^2 
\end{bmatrix},
\label{eq:subQ_matrix_u}
\end{equation}
where $ \bm{q}_{\Delta\bm{\mu}} = [q_d \quad q_{\varphi} \quad q_{\theta}]^{\text{T}} \in \mathbb{R}^{3\times 1} $ and the square root of each entry in the vector denotes the acceleration of corresponding structural parameter. The sub-matrices related to $ \alpha $ and $ \phi $ are given as $ \Q^1_{\bm{\alpha}} = q_{\alpha}\I_4$ and $ \Q^1_{\bm{\phi}} = q_{\phi}\I_4$. It should be noted that the evolutions $ \Delta\alpha $ and $ \Delta\phi $ are not involved in the state space model. However, we assign small values to the variances $ q_{\alpha} $ and $ q_{\phi} $ to account for slow variations of $ \alpha $ and $ \phi $ during the propagation processes in the free space. The same variances are assumed for different polarimetric transmission coefficients. The selection and tuning process of the noise variance are very important especially for the narrowband case, because the orthogonality is not tightly held between close-by MPCs \cite{7239623}. Small variance may lead to smooth but slow tracking, and some small movements might be missed. Large variance enables quick response to non-smooth movements like sharp turns, but leads to high risk of phase slip. Hence, a trade-off is needed. Here, we follow the guideline that the value of $ \sqrt{q_{[\,\bm{\cdot}\,]}}  $ should be in the same order as the maximum acceleration magnitude \cite{BarShalom2002EstimationTracking}. The extension of the matrices (\ref{eq:transf_matrix}) and (\ref{eq:Q_matrix}) to the multipath case is done with a Kronecker operation as $ \F = \F^1 \otimes \I_{L_n} $ and $ \Q = \Q^1 \otimes \I_{L_n} $ \cite{SalmiTSP2009}.

The corresponding linearized measurement model, which describes the non-linear mapping from MPC parameters to channel measurement, is defined as
\begin{align}
\bm{y}_n = \bm{s}(\bm{x}_{n}) + \bm{r}_n,
\label{eq:MeaModel}
\end{align}
where $ \bm{r}_n $ contains the measurement noise with covariance matrix $ \R_n $ defined in Section~\ref{subsec_NoiseVariance} and $ \bm{s}(\bm{x}_{n}) $ represents the non-linear mapping from the MPC parameters to the specular observation vector described in (\ref{eq:nonlinear_mapping}). The first-order Taylor series approximation can be used for linearizing the model $ \bm{s}(\bm{x}_{n}) $, and the linearized measurement matrix is represented with the Jacobian matrix $ \J_n $ described in (\ref{eq:Jacobian}).

\subsection{MPC Parameters Tracking Using an EKF}
\label{s_ekf}

The MPC parameters are tracked using an EKF similar to \cite{SalmiTSP2009} starting from time $n = 2$, where the state vector at time $ n = 1 $ and the estimated noise covariance matrix $ \hat{\R}_n $ are provided by the RIMAX algorithm (see Section\,\ref{s_initialize}). The filtered posterior state vector $ \hat{\bm{x}}_{n} $ is given by
\begin{equation}
\hat{\bm{x}}_n = [\hat{\bm{\mu}}_n^\text{T} \quad \Delta\hat{\bm{\mu}}_n^\text{T} \quad \hat{\bm{\alpha}}_n^\text{T} \quad \hat{\bm{\phi}}_n^\text{T}]^\text{T} \in \mathbb{R}^{14 K_n\times 1},
\label{eq:PosteriorStateVec_est}
\end{equation}
where the sub-vectors are given as
\begin{align}
\hat{\bm{\mu}}_n &= [\hat{\bm{d}}_n \quad \hat{\bm{\varphi}}_n \quad \hat{\bm{\theta}}_n]^{\text{T}} \in \mathbb{R}^{3 K_n\times 1},\\[0.3em]
\hat{\bm{\alpha}}_n &= [\hat{\bm{\alpha}}_{\text{HH},n} \quad \hat{\bm{\alpha}}_{\text{HV},n} \quad \hat{\bm{\alpha}}_{\text{VH},n} \quad \hat{\bm{\alpha}}_{\text{VV},n}]^{\text{T}} \in \mathbb{R}^{4 K_n\times 1}, \\[0.3em]
\hat{\bm{\phi}}_n &= [\hat{\bm{\phi}}_{\text{HH},n} \quad \hat{\bm{\phi}}_{\text{HV},n} \quad \hat{\bm{\phi}}_{\text{VH},n} \quad \hat{\bm{\phi}}_{\text{VV},n}]^{\text{T}} \in \mathbb{R}^{4 K_n\times 1}.
\label{eq:subvectors_est}
\end{align}

\subsubsection{Prediction Step}
The prior state vector $ \hat{\bm{x}}^{-}_{n} $ and the prior filter error covariances matrix $ \P^{-}_{n} $ given measurements up until time $ n-1 $, are respectively given by
\begin{align}
\hat{\bm{x}}^{-}_{n} &= \F \hat{\bm{x}}_{n-1},\\[0.3em]
\P^{-}_{n} &= \F \P_{n-1}\F^{\text{T}} + \Q.
\end{align}

\subsubsection{Measurement Update Step} 
\label{EKF_mea_update}
The measurement at time $n$ is used to update the predicted state vector $ \hat{\bm{x}}^{-}_{n} $ and the corresponding matrix $ \P^{-}_{n} $, resulting into the posterior covariance matrix $ \P_{n} $ and posterior state vector $ \hat{\bm{x}}_{n} $, obtained by 
\begin{align}
\P_{n} &= (\I_{14 K_n} + \K_n\D_n)\P^{-}_{n},\\[0.3em]
\Delta\hat{\bm{x}}_{n} &= \P_{n}\bm{q}_n,\\[0.3em]
\hat{\bm{x}}_{n} &= \hat{\bm{x}}^{-}_{n} + \Delta\hat{\bm{x}}_{n}\, ,
\end{align}
where the Kalman gain matrix $ \K_n $ is formulated as
\begin{align}
\K_n = \P^{-}_{n}\D^{\text{H}}_n(\D_n\P^{-}_{n}\D^{\text{H}}_n +\hat{\R}_n )^{-1},
\label{eq:KalmanGain}
\end{align}
and $ \bm{q}_n \in \mathbb{R}^{14 K_n\times 1} $ is the score function and $ \D_n \in \mathbb{R}^{14 K_n\times 14 K_n} $ represents the Fisher information matrix, which are the first-order and the second-order partial derivatives of the negative log-likelihood function, respectively. The score function $ \bm{q}_n $ and the Fisher information matrix $ \D_n $ are given by
\begin{align}
\bm{q}_n
&= 2\Re\left\lbrace \J_n^{\text{H}}\hat{\R}_n^{-1}(\bm{y}_n-\bm{s}(\hat{\bm{x}}^{-}_{n}))\right\rbrace,\\[0.3em]
\D_n
&= 2\Re\left\lbrace \J_n^{\text{H}}\hat{\R}_n^{-1}\J_n\right\rbrace\, ,
\end{align}
where the Jacobian matrix $ \J_n \in \mathbb{C}^{ N_\text{f} N_{\text{Rx}}\times 14 K_n} $ represents the the first-order partial derivatives of the linearized signal vector $ \bm{s}(\hat{\bm{x}}^{-}_{n})$, i.e.,
\begin{equation}
\J_n = \dfrac{\partial \bm{s}(\hat{\bm{x}}^{-}_{n})}{\partial (\hat{\bm{x}}^{-}_{n})^\text{T}}.
\label{eq:Jacobian}
\end{equation}

\subsubsection{State Dimension Adjustment}
\label{s_sda}

During the channel measurements, the number of tracked MPCs may vary over time. The birth and death processes of MPCs are assumed to be statistically independent and therefore the state dimension adjustment is performed alongside with the EKF.

\paragraph{Birth of MPC} Potentially new MPCs are detected in the initialization process using the SAGE algorithm as described in Section~\ref{s_initialize}, and the estimated covariance matrix $ \hat{\R}_n $ at time $n$ is used to estimate the complex weight with (\ref{eq:gammaestre}) given below.


\paragraph{Death of MPC} The posterior covariance matrix $ \P_{n} $ comprises the uncertainties of the state vector after update with measurement. Using the contained variances of the complex MPC weights, a reliability measure of a MPC is calculated and used to adjust the dimension of the state space vector $\hat{\bm{x}}_{n}$, i.e., to control the death of MPCs. At first, the SINR of each MPC \cite{RichterPhD2005} is calculated, i.e.,
\begin{equation}
\text{SINR}_{k,n} = \sum_{\text{p}}\dfrac{\mid \hat{\gamma}^\text{p}_{k,n} \mid^2}{v^\text{p}_{k,n}}\, ,
\label{eq:relativevar}
\end{equation}
where $ |\hat{\gamma}^\text{p}_{k,n}|$ is the magnitude of the estimated MPC weight for polarization $ \text{p}\in\lbrace\text{{\footnotesize HH}}, \text{{\footnotesize HV}}, \text{{\footnotesize VH}}, \text{{\footnotesize VV}}\rbrace $ and $ v^\text{p}_{k,n} $ is the estimated variance of MPC weight. A MPC is considered as unreliable if the SINR is below a predefined detection threshold $\varepsilon_{\text{r}}$, i.e., $\text{SINR}_{k,n} < \varepsilon_{\text{r}}$, and therefore it is removed from the state vector. Hence, the MPC lifetime is here defined as the time duration that the SINR of a MPC is above a given threshold, which to some extended is geometry-independent. An intuitive choice for the detection threshold $\varepsilon_{\text{r}}$ is $0\,$dB.

\subsubsection{Reinitalization of complex weights}

Even though the complex weights of the MPCs are assumed to vary only slowly in free space propagation, larger changes are expected due to small scale fading in the propagation processes, e.g., reflection, scattering, etc. Since the evolution of complex amplitudes $ \bm{\gamma}_{k,n} $ is not included into the prediction model, a reinitialization of complex weights $ \bm{\gamma}_{k,n} $ is performed to be able to follow these abrupt changes. Using the posterior MPC parameters and the estimated covariance matrix $ \hat{\R}_n $ at time $n$, the reinitialization is performed by using (\ref{eq:gammaestre}) after the mobile agent being moved a distance of about one wavelength.

\subsection{MPC Parameters and Noise Parameters Initialization with RIMAX}
\label{s_initialize}
%

Given the baseband signal $ \widetilde{\bm{y}}_n $ at time $n$, depending on the time instance, the parameters of new MPCs, i.e., $\hat{\bm{\mu}}_{k', n} $ with $k'\,\in\,{\mathcal{K}'_{n}} = \lbrace 1, \dots, {K'_{n}}\rbrace  $, are either estimated for the first time using $ \widetilde{\bm{y}}_n = \bm{y}_1$ at time $n=1$, or using the residual $ \widetilde{\bm{y}}_n = \bm{y}_n - \sum_{k \in \mathcal{K}_{n-1}} \bm{s}(\hat{\bm{\mu}}_{k,n},\hat{\bm{\gamma}}_{k,n}) $, at time $n = 2,\dots,N $, where $ \bm{s}(\hat{\bm{\mu}}_{k,n},\hat{\bm{\gamma}}_{k,n}) $ represents the specular contribution of each MPC which is already inside the state vector defined in (\ref{eq:nonlinear_mapping}). After the parameters of $ K'_n $ MPCs are estimated, the state vector will contain finally the parameters of $ K_n = K'_n + K_{n-1}$ MPCs, and $ K_{-1} = 0 $. The estimation of each new MPC is discussed in the following section. 


\subsubsection{Successive cancellation of MPCs}\label{p_initiaMPC}
The estimation starts by using the space-alternating generalized expectation-maximization (SAGE) algorithm that is based on successive cancellation of MPCs \cite{BernardFleury_Journal1999}. At first, an initial estimate of the $ k' $th new MPC's parameters $\hat{\bm{\mu}}_{k', n} $ is found by searching for the maximum of the power spectrum of $ \widetilde{\bm{y}}_n^{-} $, with $ \widetilde{\bm{y}}_n^{-} = \widetilde{\bm{y}}_n $ when $ k' = 1 $, i.e.,\footnote{$ \widetilde{\bm{y}}_n $ is reshaped to a matrix with dimension $ N_{\text{Rx}} \times N_\text{f} $ before used in (\ref{eq:pathest}).}  
\begin{equation}
\{\hat{n}_a',\hat{n}_e',\hat{i}\}=\arg\max_{n_a',n_e',i} |\bm{b}^{\text{Rx}}(\varphi_{\text{s},n_a'},\theta_{\text{s},n_e'})\widetilde{\bm{y}}_n^{-}\bm{a}_{\text{f},i}^*|, 
\label{eq:pathest}
\end{equation}
where $\hat{\bm{\mu}}_{k',n} =  [d_{\hat{i}}',\varphi_{\text{s}, \hat{n}_a'} \theta_{\text{s},\hat{n}_e'}]^{\text{T}}$, and $\bm{a}_{\text{f},i}$ is the $ i $th column of $\mathbf{A}_\text{f}$ in (\ref{eq:ComplexShiftingMatrix}). $ \varphi_{\text{s},n_{\text{a}}'} $ and $ \theta_{\text{s},n_{\text{e}}'} $ are the azimuth and elevation angles after interpolation, with $ n_{\text{a}}' = 1,\cdots, N_{\text{a}}'$ and $ n_{\text{e}}' = 1,\cdots, N_{\text{e}}'$. $ N_{\text{a}}' $ and $ N_{\text{e}}' $ denote the number of azimuth, elevation angular samples after interpolation. The sub-vectors are given as
\begin{equation}
\bm{b}^{(\text{Rx})}(\varphi_{\text{s},n_{\text{a}}'},\theta_{\text{s},n_{\text{e}}'}) = \bm{b}^{(\text{Rx})}_{\text{H}}(\varphi_{\text{s},n_{\text{a}}'},\theta_{\text{s},n_{\text{e}}'}) + \bm{b}^{(\text{Rx})}_{\text{V}}(\varphi_{\text{s},n_{\text{a}}'},\theta_{\text{s},n_{\text{e}}'}).
\label{eq:Intepolated_BP_Matrix_subvector}
\end{equation}
The vector $ \bm{b}_{(\text{H}/\text{V})}^{\text{Rx}} \in \mathbb{C}^{ N_{\text{Rx}} \times 1 } $ represents the projection from $ (\varphi_{\text{s},n_{\text{a}}'},\theta_{\text{s},n_{\text{e}}'}) $ to the array response by using the effective aperture distribution function (EADF). At the transmit side, the antenna response is denoted by a scalar $ b_{(\text{H}/\text{V})}^{\text{Tx}} $ due to a single antenna being used. The EADF performs efficient interpolation of the measured beam pattern via a two-dimensional discrete Fourier transform to obtain antenna responses of arbitrary azimuth and elevation angles that are off the sampling grid. The reader is referred to \cite{Landmann_EADF_2004} for more details regarding the EADF formulation. For interpolation in the delay/distance domain, the complex shifting matrix
\begin{align} 
\A_{\text{f}} = 
\begin{bmatrix}
\text{e}^{-j 2\pi(-\frac{N_\text{f}-1}{2}) f_1'}& \dots & \text{e}^{-j 2\pi(-\frac{N_\text{f}-1}{2}) f_{N_\text{f}'}'}\\
\vdots &  & \vdots\\[0.3em]
\text{e}^{-j 2\pi(+\frac{N_\text{f}-1}{2}) f_1'} & \dots & \text{e}^{-j 2\pi(+\frac{N_\text{f}-1}{2}) f_{N_\text{f}'}'}
\end{bmatrix}\, \in \mathbb{C}^{N_\text{f} \times N_\text{f}'}
\label{eq:ComplexShiftingMatrix}
\end{align}
is applied for an increased number of frequency points $ f_{i}' = \frac{i}{N_\text{f}'}$ with $ i = 1, \dots, N_\text{f}' $, where $N_\text{f}'$ denotes the number of frequency samples after interpolation. The corresponding distance samples are given as $ d_i' = \frac{c  N_\text{f}}{B_{\text{w}}}f_{i}' $. 
The estimate of the corresponding complex weight $ \hat{\bm{\gamma}}_{k',n} $ is given in two forms. If there is no estimate of noise covariance matrix, $ \hat{\R}_n = \bm{I}_{N_\text{f} N_{\text{Rx}}}$ is then assumed and $ \hat{\bm{\gamma}}_{k',n} $ is given in a least square form, i.e.,
\begin{equation}
\hat{\bm{\gamma}}_{k',n} = (\B^{\text{H}}(\hat{\bm{\mu}}_{k',n}) \B(\hat{\bm{\mu}}_{k',n}))^{-1} \B^{\text{H}}(\hat{\bm{\mu}}_{k',n})\widetilde{\bm{y}}_n^{-}.
\label{eq:gammaest}
\end{equation}
Otherwise, with the estimated $ \hat{\R}_n  $, $ \hat{\bm{\gamma}}_{k',n} $ is given in a weighted least square form, i.e.,
\begin{align}
&\hat{\bm{\gamma}}_{k',n} = (\B^{\text{H}}(\hat{\bm{\mu}}_{k',n}) \hat{\R}_n^{-1} \B(\hat{\bm{\mu}}_{k',n}))^{-1} \B^{\text{H}}(\hat{\bm{\mu}}_{k',n})\hat{\R}_n^{-1}\widetilde{\bm{y}}_n^{-},
\label{eq:gammaestre}
\end{align} 
where the matrix valued function $ \B(\hat{\bm{\mu}}_{k',n}) \in \mathbb{C}^{N_\text{f} N_{\text{Rx}}\times 4} $ accounts for the structure of the radio channel of four polarimetric transmissions and is defined as 
\begin{align}
&\B(\hat{\bm{\mu}}_{k',n}) = [\bm{b}_{\text{H}}^{\text{Rx}}\lozenge b_{\text{H}}^{\text{Tx}}\lozenge\bm{b}_{\text{f}}\quad \bm{b}_{\text{V}}^{\text{Rx}}\lozenge b_{\text{H}}^{\text{Tx}}\lozenge\bm{b}_{\text{f}}\quad \nonumber \\ & \qquad \qquad \quad \qquad \quad \cdots \quad \bm{b}_{\text{H}}^{\text{Rx}}\lozenge b_{\text{V}}^{\text{Tx}}\lozenge\bm{b}_{\text{f}}\quad \bm{b}_{\text{V}}^{\text{Rx}}\lozenge b_{\text{V}}^{\text{Tx}}\lozenge\bm{b}_{\text{f}}].
\label{eq:MatrixValuedFunc}
\end{align}
The vector $ \bm{b}_{\text{f}} \in \mathbb{C}^{ N_{\text{f}} \times 1 }  $ accounts for the system frequency response by using the complex shifting matrix defined in (\ref{eq:ComplexShiftingMatrix}). The detailed formulation of the matrix valued function (\ref{eq:MatrixValuedFunc}) can be found in \cite{RichterPhD2005}. The non-linear mapping from the estimated parameters of $k'$th MPC to the specular observation vector\footnote{Given an estimated state vector $ \hat{\bm{x}}_{n} $ of $ K_n $ MPCs, the specular observation vector is given as $ \bm{s}(\hat{\bm{x}}_{n}) = \sum_{k \in \mathcal{K}_{n}} \bm{s}(\hat{\bm{\mu}}_{k,n},\hat{\bm{\gamma}}_{k,n}) $} is given as 
\begin{equation}
\bm{s}(\hat{\bm{\mu}}_{k',n},\hat{\bm{\gamma}}_{k',n}) = \B(\hat{\bm{\mu}}_{k',n})\hat{\bm{\gamma}}_{k',n}.
\label{eq:nonlinear_mapping}
\end{equation}
The estimated specular component of the $ k' $th MPC, together with the components of $ k'-1 $ previously initialized new MPCs indexed with $ j $ are then subtracted from the channel observation, i.e., the residual is updated to
\begin{equation}
\widetilde{\bm{y}}_n^{-} = \widetilde{\bm{y}}_{n} - \bm{s}(\hat{\bm{\mu}}_{k',n},\hat{\bm{\gamma}}_{k',n}) - \sum_{j = 1}^{k'-1} \bm{s}(\hat{\bm{\mu}}_{j,n},\hat{\bm{\gamma}}_{j,n}).
\label{eq:pathTakeaway}
\end{equation}
A new MPC is further initialized from the updated residual $ \widetilde{\bm{y}}_n^{-} $ only if two constraints are both met: (i) The maximum allowed number of MPCs that are estimated/tracked simultaneously $ K_{\text{max}} $ is limited, so $ K_{n-1} + k' < K_{\text{max}} $; (ii) The ratio between the energy sum of the estimated MPCs, with each denoted as $ p_{\text{sp},k/j,n} $, over the full signal energy $ p_{n} $ at time $ n $, i.e., $ \beta_{n} = \sum_{k \in \mathcal{K}_{n-1}}\frac{p_{\text{sp},k,n}}{p_{n}} + \sum_{j = 1}^{k'}\frac{p_{\text{sp},j,n}}{p_{n}} $, should be smaller than the maximum allowed ratio denoted as $ \beta_{\text{max}} $, i.e., $ \beta_{n} < \beta_{\text{max}} $. This is to control the model complexity and reduce the interference between coherent MPCs. The same procedure from (\ref{eq:pathest}) to (\ref{eq:pathTakeaway}) is repeated until the parameters of $ K'_n $ MPCs are estimated and added to the state vector and $ K_n = K'_n + K_{n-1}$. 

After subtracting the contribution of the $ K_n $ estimated MPCs, the residual $ \widetilde{\bm{y}}_n^{-} $ is used to estimate the noise standard deviation and DMC parameters $\hat{\bm{x}}_{\text{noise,n}} = [\hat{\bm{x}}_{\text{dmc},n}^{\text{T}}\quad \hat{\sigma}_{\text{w},n}]^{\text{T}}$. The initial estimates of $ \hat{\bm{x}}_{\text{dmc},n} $ is computed from the averaged power delay profile over $ N_{\text{Rx}} $ antenna elements. The reader is referred to \cite[Section 6.1.8]{RichterPhD2005}) for detailed processing. The estimated covariance matrix $ \hat{\R}_n $ is then calculated using $ \hat{\bm{x}}_{\text{noise,n}} $ with \eqref{eq:fullCov} and \eqref{eq:TopeRf}.

\subsubsection{Refinement with RIMAX} These initial estimated state vectors of MPC parameters, noise and DMC parameters are optimized by alternating maximization of the log-likelihood function by the RIMAX algorithm, which uses Levenberg-Marquardt and ML-Gauss-Newton algorithms for optimization \cite{RichterPhD2005}. It is worth mentioning that $ K_n $ influences how much one can actually benefit from the joint optimization of all parameters in RIMAX initialization. If a nonsensical solution with a very large $ K_n $ is given in (\ref{p_initiaMPC}), the estimated parameters of specular MPCs tend to converge to local minima which are biased from the true values due to noise over-fitting after optimization iterations. In practice, the maximum $ K_n $ allowed in the RIMAX initialization should be chosen to capture all the significant specular MPCs in the propagation environment. Considering the alternating maximization to jointly estimate the MPC parameters $ \hat{\bm{x}}_n $ and the noise parameters $\hat{\bm{x}}_{\text{noise,n}} = [\hat{\bm{x}}_{\text{dmc},n}^{\text{T}}\quad \hat{\sigma}_{\text{w},n}]^{\text{T}}$ is very computationally demanding and therefore it's only applied at time $ n=1 $, but not during the subsequent tracking of the channel parameters.

\section{Localization and Mapping}
\label{s4}
Given the distance estimates $ \hat{d}_{k,n} $ from EKF, the localization problem is formulated as the joint estimation and optimization process of the inlier set $ I_{\text{inl}} $, mobile agent positions $\bm{p}_n$ and features (PA and VAs) positions $\bm{a}_{k}$ in Problem \ref{prob_optmisstoa} in (\ref{eq:optim}), which is a highly non-convex problem. To make it a better conditioned problem with reasonable complexity, we introduce two modified versions of (\ref{eq:optim}) with given assumptions and prior information. 

\subsection{Experiment \RomanNumeralCaps{1}}
\label{ExpI}
In this experiment, we assume that all mobile agent positions $\bm{p}_n$ are known, then the optimization problem of Problem \ref{prob_optmisstoa} in (\ref{eq:optim}) is reduced to
\begin{align}
\min_{I_{\text{inl},k},\bm{a}_{k} \atop \forall n}\, \sum_{k|(k,n) \in I_{\text{inl}}} (\hat{d}_{k,n}- \norm{\bm{p}_n-\bm{a}_{k}} )^2 + \sum_{k|(k,n) \in I_{\text{outl}}} C
\label{eq:optim2}
\end{align}
independently for each feature position $\bm{a}_{k}$ with $ k \in \mathcal{K}_{n} $ and $ n = 1,\dots N $, by using random sample consensus (RANSAC) \cite{Fischler:1981:RSC:358669.358692}. $ I_{\text{inl},k} $ is the inlier subset for each $ k $ and it is possible to have no inliers at some estimated feature positions, i.e., $ I_{\text{inl},k} = \emptyset $. We assume the association between an estimated MPC and a feature is consistent during the tracking process, and the corresponding feature position is fixed over time. Given a vector containing all the distance estimates of one MPC, we randomly choose a minimal set, i.e., estimates at three time instances and corresponding (known) mobile agent positions, to give an initial estimate of the feature position. Then, we extend to the full vector and determine how many of the remaining estimates agree with the estimated feature position, i.e., number of inliers. Different minimal sets give different solutions, we choose the one with the largest number of inliers. The same procedure is repeated for all the tracked MPCs. The RANSAC gives initial estimates of feature positions and inlier set, which is then followed by non-linear optimization of (\ref{eq:optim2}). Moreover, the resulting residuals $ \hat{d}_{k,n}- \norm{\bm{p}_n-\hat{\bm{a}}_{k}}$ can be used to empirically assess properties of the error distribution. 

\subsection{Experiment \RomanNumeralCaps{2}}
\label{ExpII}
In this experiment, we assume that both the mobile agent positions and feature positions are unknown. To make this estimation problem tractable, an estimate of the inlier set $ \hat{I}_{\text{inl}} $ is firstly obtained by using the groundtruth mobile agent positions in (\ref{eq:optim2}), then the estimates $ \hat{d}_{k,n} \forall~(k,n)~\in~ \hat{I}_{\text{inl}} $ are subsequently used as inputs in (\ref{eq:optim}) and Problem \ref{prob_optmisstoa} is reduced to
\begin{align}
\min_{\bm{p}_n,\bm{a}_{k} \atop \forall n, \forall k}\, \sum_{(k,n) \in \hat{I}_{\text{inl}}} (\hat{d}_{k,n}- \norm{\bm{p}_n-\bm{a}_{k}} )^2.  
\label{eq:optim3}
\end{align}

\begin{figure}[t!]
	\centering
	\includegraphics[width=0.45\textwidth]{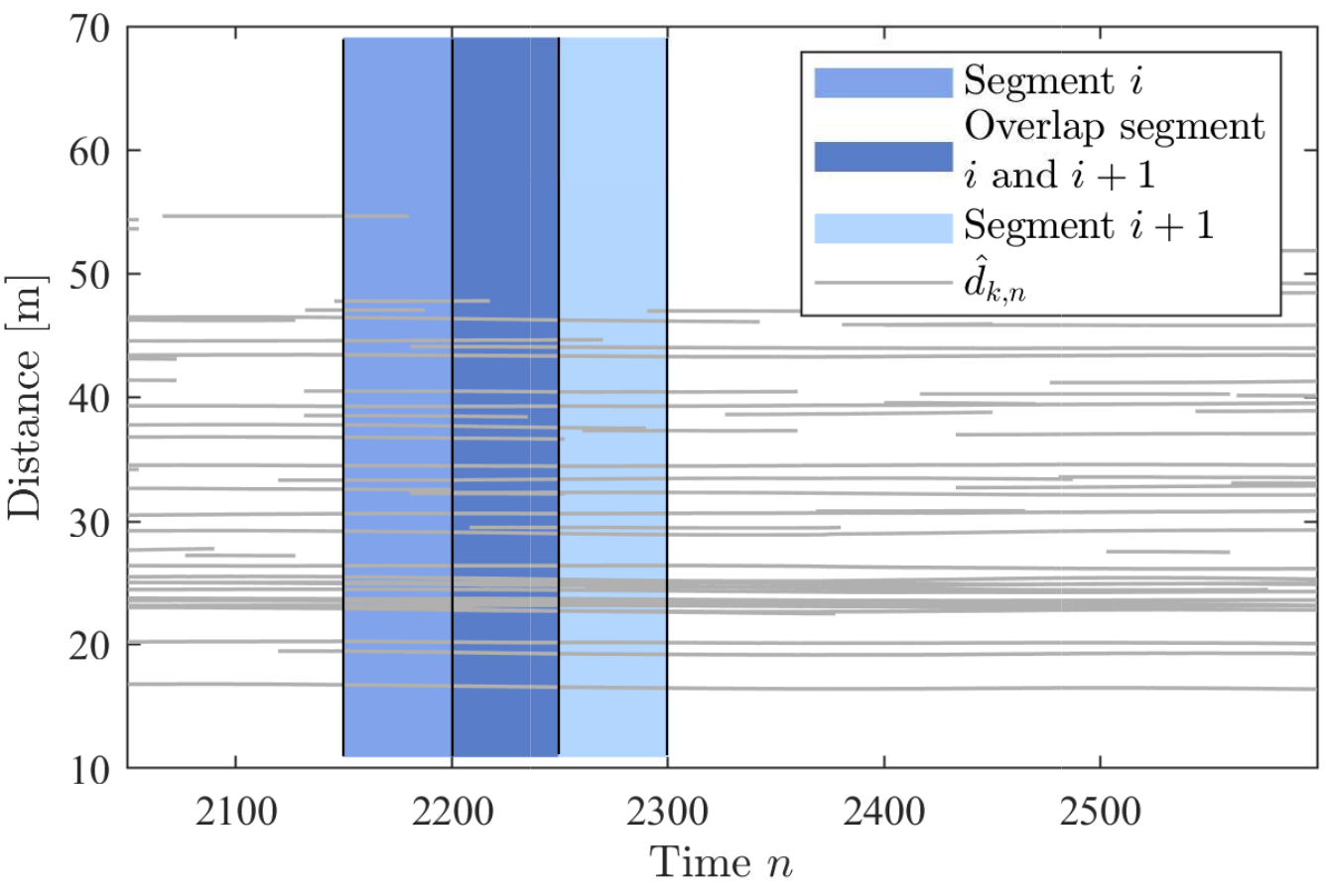}
	\caption{Depiction of how the estimated dataset from EKF being segmented before used in the localization and mapping algorithm. Each segment contains distance estimates from $ 100 $ consecutive time instances, and the overlap in-between is $ 50 $ time instances long.}
	\label{fig:segments}
	\vspace{0pt}
\end{figure}

Two assumptions are further made here. First, $ \bm{p}_n $ are assumed to be constrained to a plane (e.g.,\ $z_{n}=0$). This is a natural assumption for many problems where the mobile agent is moved approximately in a plane during the measurement. Note that it does however introduce an ambiguity in the feature positions, since we can never determine the sign of the $z-$component. Second, we assume that the mobile agent has been moved in a continuous path. Algorithms for solving Problem~\ref{prob_optmisstoa} using hypothetical and test paradigm are also presented in \cite{7760673}. In order to minimize drift and accumulation of initialization errors, we divide the whole dataset into a number of smaller segments in time (typically containing 100 time instances each). Fig.~\ref{fig:segments} shows two consecutive segments and the overlap in-between, each segment is then initialized independently. 

%


For each segment, we initialize both $ \hat{\bm{a}}_{k} $ with $ k \in \mathcal{K}_{n} $ and $ \hat{\bm{p}}_n $ using minimal solvers and RANSAC \cite{burgess-kuang-etal-sp-15} based only on the distance estimates from the EKF. In detail, we start from a minimal set which is sampled from the distance estimates to estimate the corresponding mobile agent and feature positions. Since the LOS component is visible at all time instances, we always include the PA position in the minimal sample. Using the initial positions we can minimally trilaterate mobile agent and feature positions at other time instances, and count how many inliers we get for this initial estimate. The steps above are repeated and we choose the solution with the highest number of inliers. In minimal trilateration, two possible solutions are provided for each estimated position due to the ambiguity of the $z$-component. For the mobile agent position, with the assumption of the trajectory in the plane $z_{n}=0$, we always choose the solution with the smallest $|\hat{z}_{n}|$. For the feature positions, the two solutions correspond to the two different signs of $|\hat{z}_{k}|$. Since this ambiguity can never be resolved, we consistently choose the solution with e.g.,\ positive $\hat{z}_{k}$, without any loss of generality. To sum up, the RANSAC procedure provides an initial estimate of the mobile agent positions in the segment, as well as the feature positions and an estimate of the inlier set. The solution of (\ref{eq:optim}) is then refined by using a Newton method.   

The estimates for each segment are given in its own coordinate system. However, we need the whole solution to be in the same coordinate system. If we choose the segments so that the reconstructed mobile agent's MPC has an overlap in-between segments, we can use the overlapping mobile agent positions to register the different reconstructions. This is simply done in a least squares way by applying rotation and translation operations. After registration of all segments into one coordinate system, the mean values over all individual estimates are calculated for overlapping mobile agent positions and feature positions. We can then also do a final non-linear optimization of all estimated positions over all inlier data.


%

\section{Evaluations and Results}
\label{s5}
To analyze the performance, the proposed framework is applied to both real and synthetic channel measurements, and the mobile agent is equipped with one single antenna in both cases. Besides, the results are presented in two aspects: (i) the MPC parameter estimation and tracking results, and comprehensive statistical analysis of the MPC dynamic behaviors, (ii) the evaluation of two localization and mapping experiments presented in Section~\ref{s4} with real channel measurements.


\subsection{Experimental Setup}
\label{s5_1}

\begin{figure}[t!]
	\centering
	\includegraphics[width=8.3cm,height=4.05cm]{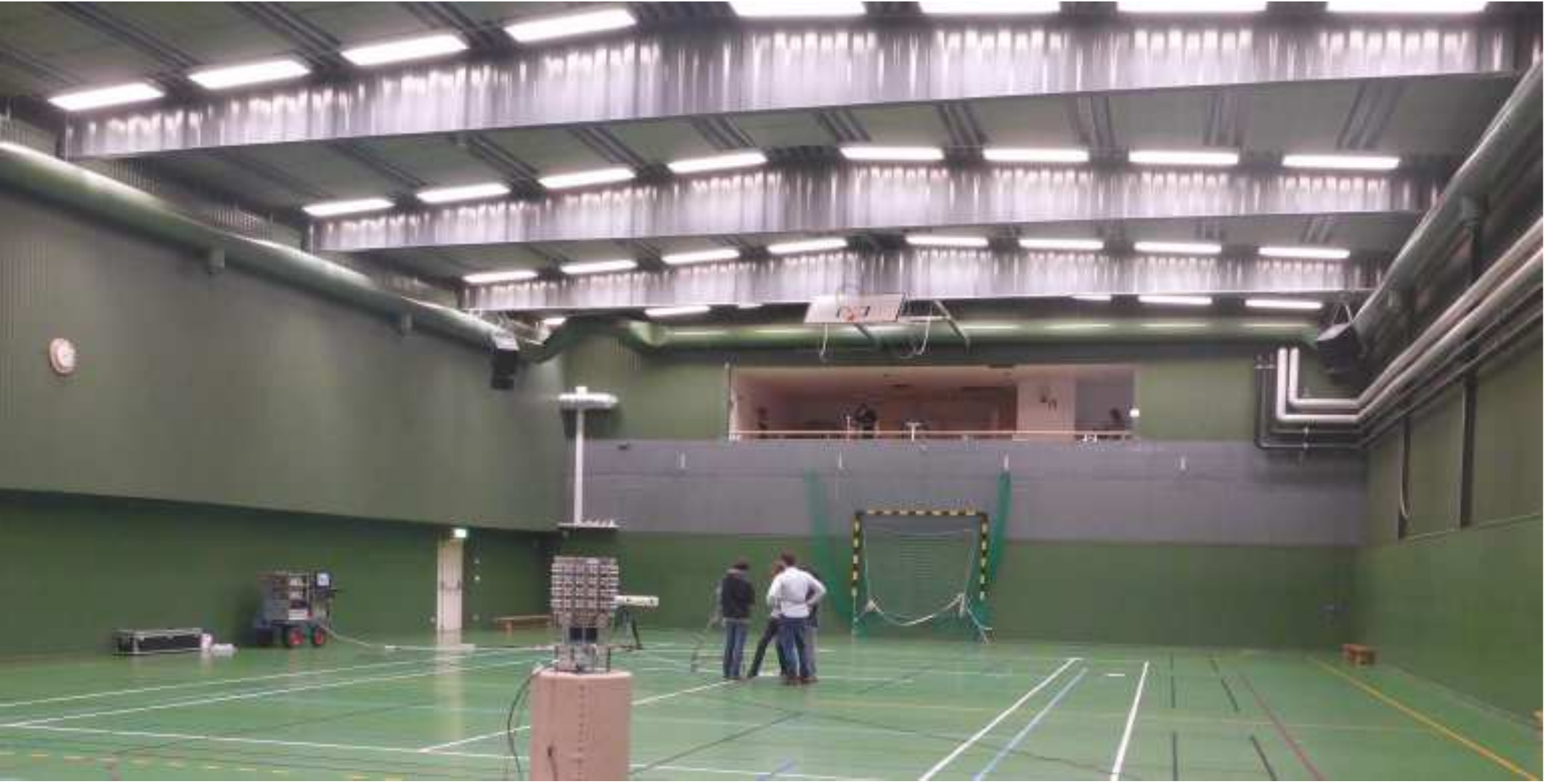}
	\caption{Overview of the measurement area in the sports hall, Medicon Village, Lund, Sweden. Room dimension is around 20\,m $ \times $ 36\,m $ \times $ 7.5\,m.}
	\label{SportHall}
	\vspace{0pt}
\end{figure}

\begin{figure}[t!]
\centering
\subfloat[]{
	\label{subfig:Antenna_Rx}
	\includegraphics[width=3.3cm,height=4.2cm]{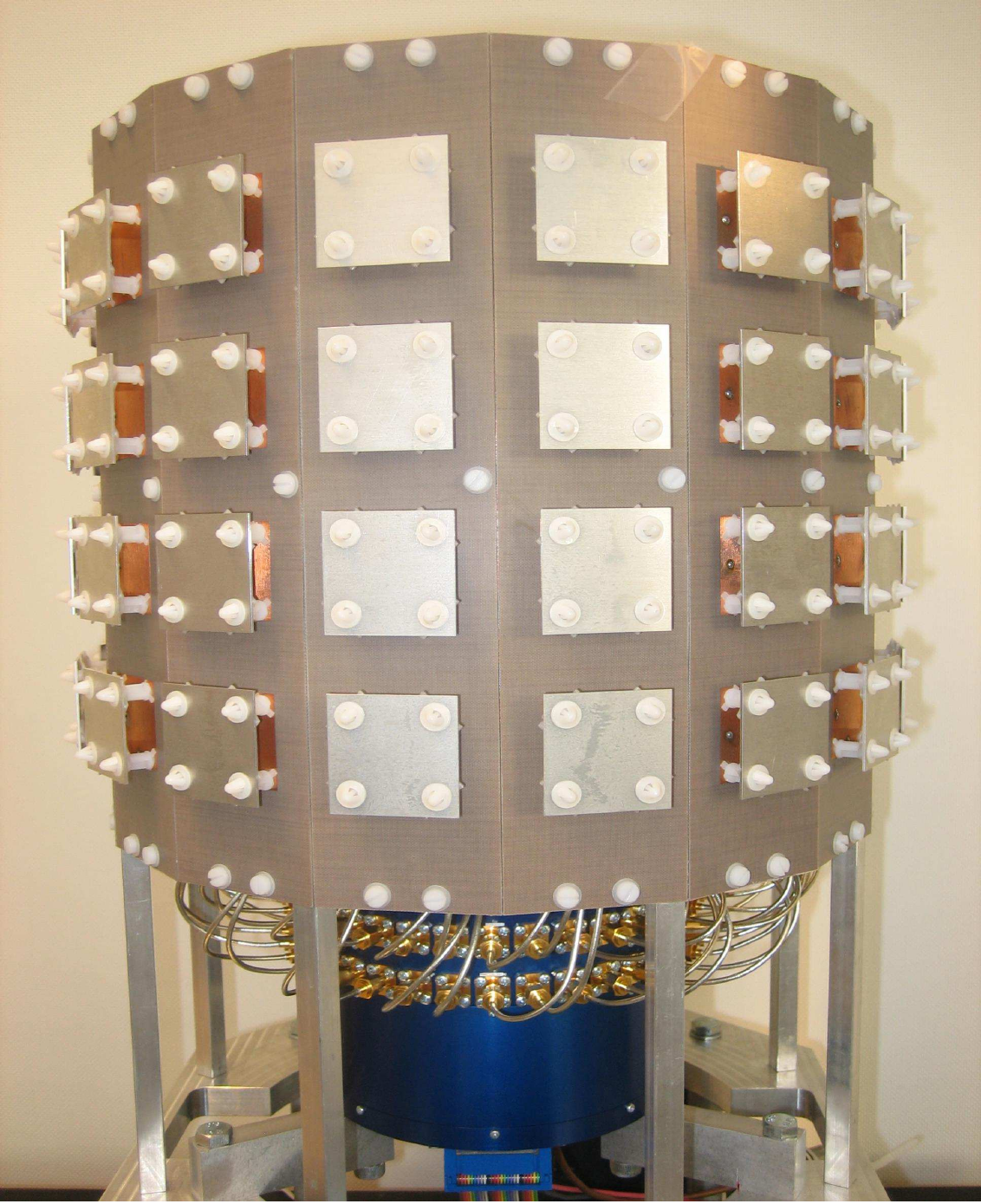}}\hspace{1 mm}
\subfloat[]{
	\label{subfig:Antenna_Tx}
	\includegraphics[width=4.1cm,height=4.2cm]{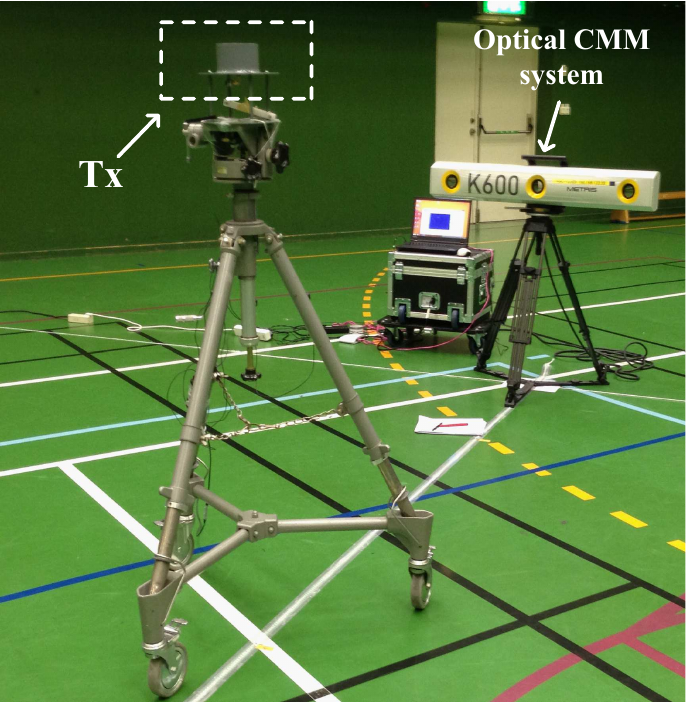}}
\caption{(a) Photo of the cylindrical antenna array. (b) The conical monopole omnidirectional antenna and the optical CMM system.}
\end{figure}

The real measurement campaign was performed in a large sports hall with the RUSK LUND channel sounder. Fig.~\ref{SportHall} shows an overview of the measurement area. A cylindrical array with 64 dual-polarized antennas (Fig.~\ref{subfig:Antenna_Rx}), i.e., 128 ports in total, is used as a static PA. The center of the array is 1.42\,m above ground. A conical monopole omnidirectional antenna (Fig.~\ref{subfig:Antenna_Tx}) is used to represent a mobile agent. The distance between the PA and the mobile agent is around 17\,m and line-of-sight (LOS) conditions apply. The transfer functions (snapshots) were recorded at a center frequency around 2.7\,GHz and with 129 frequency samples equispaced over a 40\,MHz bandwidth. To avoid large variation of path parameters, especially a possible 2$ \pi $ phase slip between two consecutive snapshots, the spatial sampling rate of the wireless channel was sufficiently high. In total, there were 6000 channel snapshots collected in 19.7\,s. The mobile agent was placed on a tripod and manually moved to write the ``\textbf{Lund}'' letters in an approximately 2\,$ \text{m}^2 $ area. Meanwhile, an optical CMM system (Fig.~\ref{subfig:Antenna_Tx}), which uses the camera technology to triangulate the positions with accuracy down to millimeter, was used to capture the mobile agent movement. The movement positions are further used as the groundtruth $ \bm{p}_{\text{true},n} $ for performance analysis. The floor plan and the zoom-in plot of the groundtruth are shown in Fig.~\ref{Theory}

A synthetic measurement dataset was generated for validating the performance of the MPC parameters' initialization. The floor plan based on the Medicon Village in Fig.~\ref{Theory} (excluding the ceiling) and a ray tracer (RT) are used to generate dispersion parameters of MPCs. During the RT simulations, the real calibration file of the cylindrical antenna array (Fig.~\ref{subfig:Antenna_Rx}) was used at the PA side and the mobile agent with a single omnidirectional antenna was assumed. DMC was also included and independently generated for each realization. The PA was kept static at the location which is the same as the real measurement setting. Meanwhile, the groundtruth coordinates of mobile agent at the first 100 time instances $ n $ from the optical system were used to synthetically generate 100 independent channel realizations. The energy ratio $ \beta_{n} $ is around 50\% for each realization. The number of MPCs was restricted to $ L_n = 6 $ for each realization, including the LOS and the first order reflection paths from surrounding walls and ground.

\subsection{Evaluation of the Channel Estimation Algorithm}
\label{s5_2}


\subsubsection{MPC Initialization Performance}
The RIMAX was applied to each synthetic channel realization independently. For consistent evaluation of the estimation errors between the reference state vector $ \bm{x}_n $ from RT and the estimated state vectors $ \hat{\bm{x}}_n $, the optimal sub-pattern assignment (OSPA) metric \cite{4567674} was applied here. For the case $ K_n \geq L_n $, it is defined as 
\begin{align}
& d_{\text{ospa}}(\hat{\bm{x}}_n,\bm{x}_n) = \biggl[ \dfrac{1}{K_n} \biggl( \min_{\pi \in \prod_{K_n}}\,\sum_{l=1}^{L_n} \left[  d^{(d_\text{c})}\left( d_{l,n},\hat{d}_{\pi_l,n}\right)\right]^{p_{\text{o}}} \biggr. \nonumber \\
& \qquad\qquad\qquad\quad \biggl. + \quad d_{\text{c}}^{p_{\text{o}}}\left(K_n-L_n \right) \biggl) \biggr]^{\frac{1}{p_{\text{o}}}},
\label{eq:ospa}
\end{align}
where $ \prod_{k} $ denotes the set of permutations on $ \lbrace 1,\cdots,k\rbrace $ and $ k \leq K_n $. $ d(\cdot,\cdot) $ represents the Euclidean metric and the function $ d^{(d_{\text{c}})}\left( \cdot,\cdot\right) = \min (d_{\text{c}}, d(\cdot,\cdot)) $. Besides, we have the cut-off parameter of distance $ d_{\text{c}} = 1 $\,m and order parameter $ p_{\text{o}} = 1 $.

\begin{figure}[t!]
\centering
\subfloat[]{
	\label{RIMAX_MPC_NUMBER}
	\includegraphics[width=0.45\textwidth]{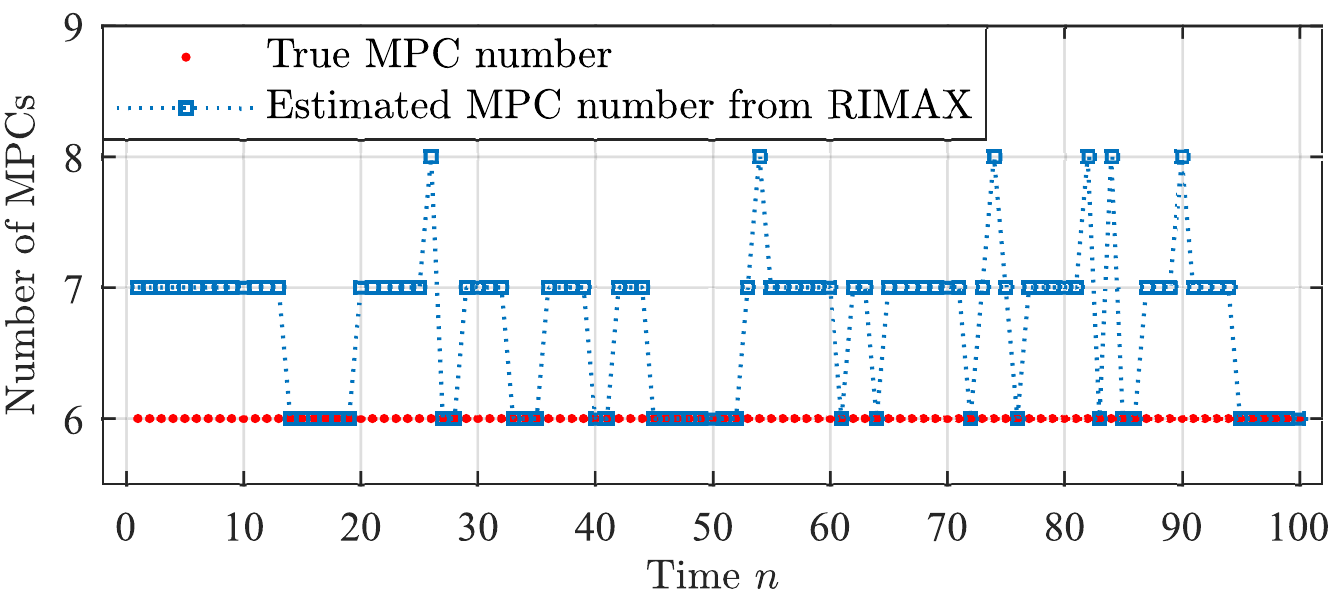}}
	\hspace{2mm}
\subfloat[]{
	\label{OSPA_RIMAX_converage}
	\includegraphics[width=0.45\textwidth]{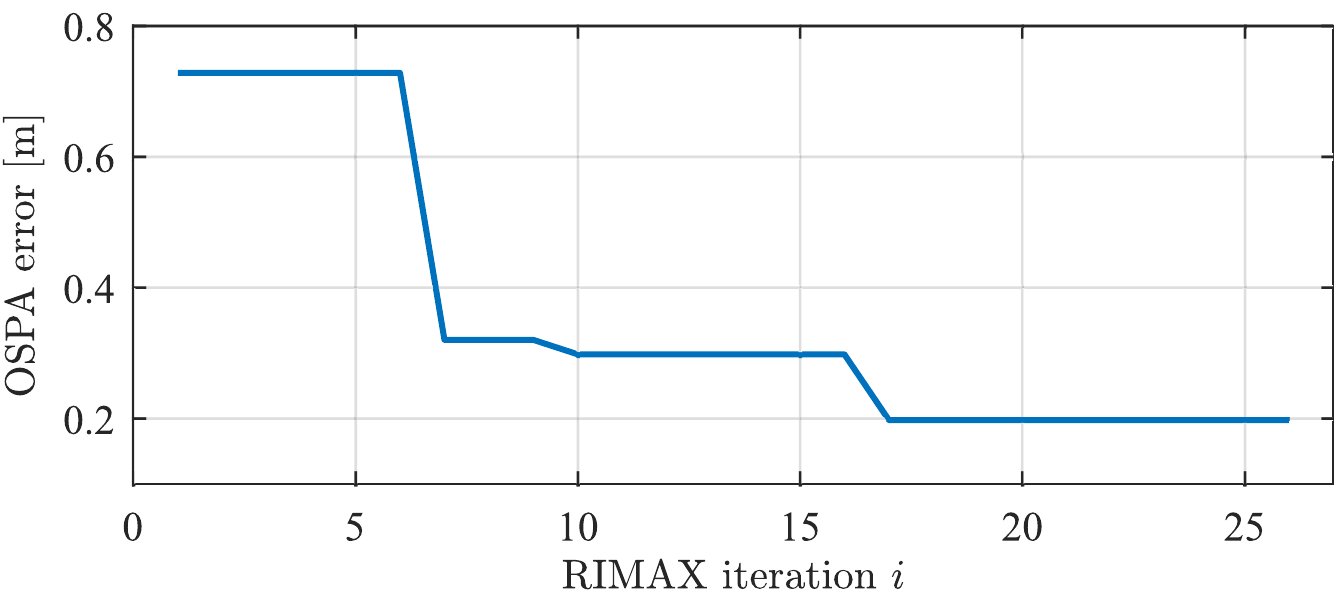}} 
\caption{(a) The number of MPCs estimated using RIMAX $ K_n $, compared with true number of MPC $ L_n $ for each channel simulation. (b) The convergence of the OSPA error versus the number of iterations of the joint optimization in RIMAX algorithm for the the synthetic channel realization at time $n = 1$.}
\label{kohler} 
\end{figure}

With the maximum allowed power ratio $ \beta_{\text{max}} $ set to 55\,$\%$, overestimation happened as expected after the SAGE step, around 20 MPCs are detected at each time $ n $. After RIMAX iterations and state dimension adjustment, clutter components are enormously suppressed from the initial state vector, only one or two clutter components remain for some time instances, as shown in Fig.~\ref{RIMAX_MPC_NUMBER}. For the synthetic realization at time $ n=1 $ as shown in Fig.~\ref{OSPA_RIMAX_converage}, the estimated state order $ K_n $ is reduced from 20 to 7 with the RIMAX iterations and dimension adjustment, meanwhile the OSPA error decreases from around 0.73\,m to 0.19\,m. The RIMAX was applied independently to the 100 synthetic channel realizations, and the mean OSPA error over 100 simulations is 0.172\,m.

\subsubsection{MPC Tracking Results}
\begin{figure}[t!]
	\centering
	\includegraphics[width=0.45\textwidth]{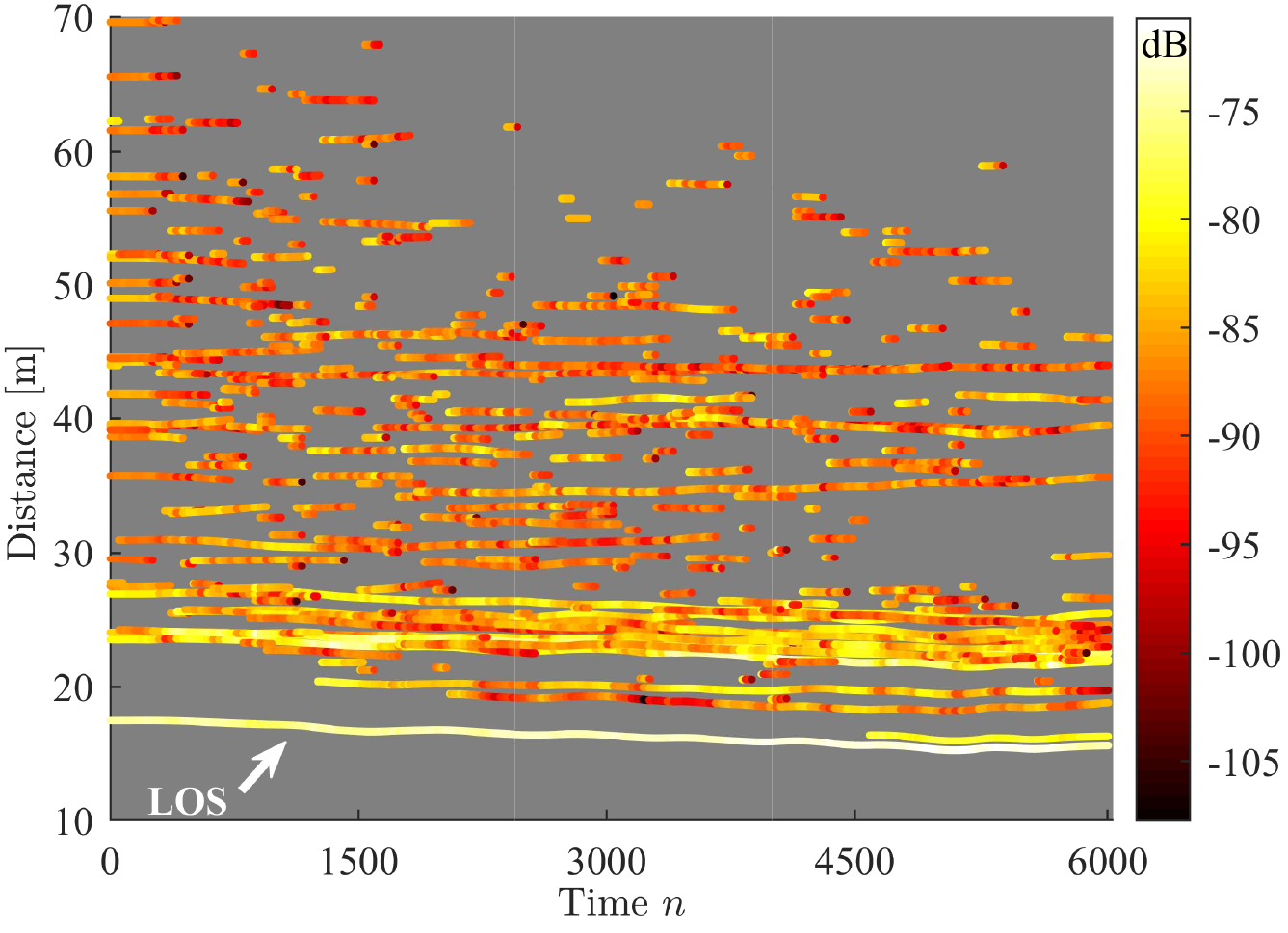}
	\caption{The tracked propagation distances of MPCs $ \hat{d}_{k,n} $ over measurement time, with the color indicating the power in dB scale.}
	\label{MPCdistance}
	\vspace{0pt}
\end{figure}

\begin{figure}[t!]
\centering
\subfloat[]{
	\label{subfig:3D_topview}
	\includegraphics[width=4cm,height=4.2cm]{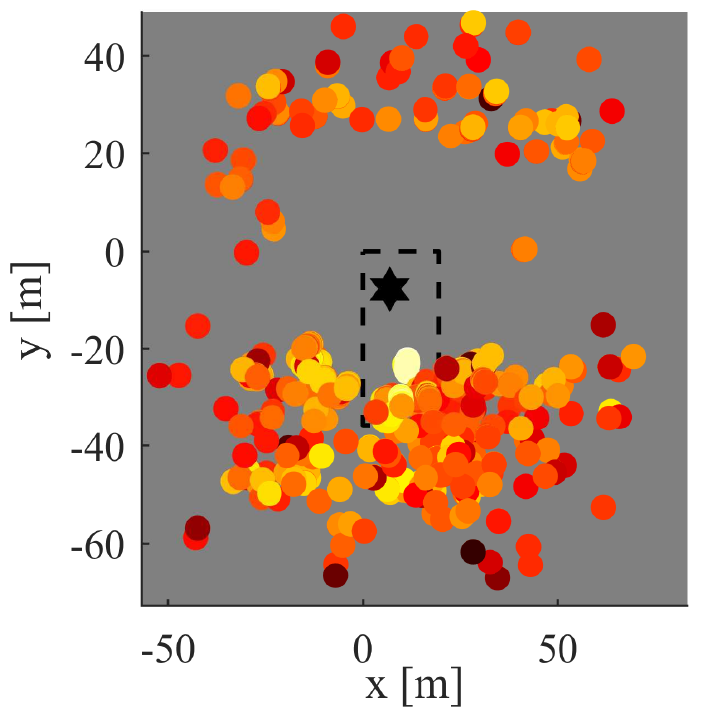}}
\subfloat[]{
	\label{subfig:3D_sideview}
	\includegraphics[width=4.8cm,height=4.2cm]{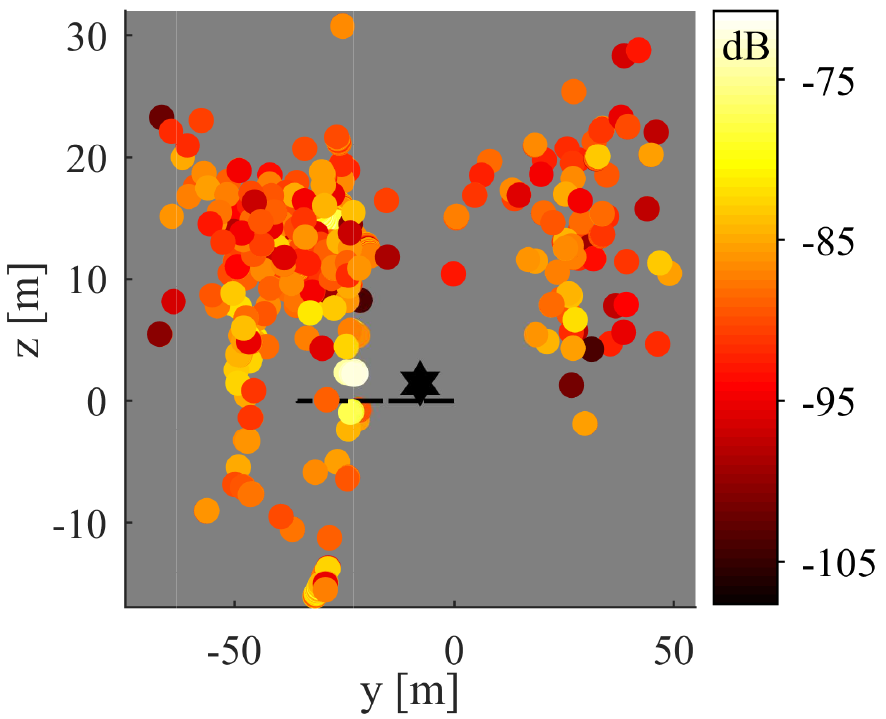}}
\caption{3D plot of the tracked MPCs based on the estimated distances $ \hat{d}_{k,n} $ and azimuth/elevation AoAs $ (\hat{\varphi}_{k,n},\hat{\theta}_{k,n}) $. Black dashed line denotes the room geometry and the hexagram represents the location of BS. The top-view plot (a) shows how tracked MPCs are distributed in the azimuth plane. The side-view plot (b) shows the vertical distribution.}
\label{3Dview}
\end{figure}

Table \ref{table_ParaEKF} summarizes the parameters initialized in the EKF estimation for the ``\textbf{Lund}'' measurement, where the noise and DMC parameters are estimated at every $5$th time instances and reinitialization of the complex amplitude $ \bm{\gamma}_{k,n} $ is performed every $36$th time instances. These values are adapted to movement of the mobile agent. Fig.~\ref{MPCdistance} shows the tracked propagation distances of MPCs over measurement time from the EKF implementation. It is observed that the LOS component with the distance around 17\,m is tracked steadily since the beginning. About 2\,m apart from the LOS is the ground reflection path which is tracked shortly in the end. Besides, many other MPCs with long lifetimes are observed in the range of 20-50\,m propagation distance. The spatial distribution of the tracked MPCs are further given in Fig.~\ref{3Dview}. The MPCs are plotted in a 3D Cartesian coordinate system based on the estimates of distances $ \hat{d}_{k,n} $ and azimuth/elevation AoAs $ (\hat{\varphi}^{(\text{Rx})}_{k,n},\hat{\theta}^{(\text{Rx})}_{k,n}) $ without considering the path interaction order. The top view (Fig.~\ref{subfig:3D_topview}) shows that the tracked MPCs are distributed over the entire azimuth domain and paths are intensively detected in the similar direction as the LOS component. From the vertical distribution (Fig.~\ref{subfig:3D_sideview}), a few paths are observed from the ground or at similar height as the PA, while the most of the estimated paths are from the complex ceiling structure of the room, e.g., the metal beams of the ceiling in Fig.~\ref{SportHall}. Those complex room structures brought additional uncertainties to the distance estimates. Moreover, the similar behaviour of the long-tracked MPCs in the angular domain may become a challenge for 3D localization, for which the MPCs with sparse angles are preferred. However, it is interesting to see the performance in the real but non-ideal case. 

\begin{table*}[t]  
\renewcommand*{\arraystretch}{1.4}
\centering
\caption{The parameters used in the EKF estimation for the ``\textbf{Lund}'' measurement.}
\label{table_ParaEKF}
\begin{tabular}{
|@{\hspace{.9mm}}c@{\hspace{.9mm}}	
 |@{\hspace{.9mm}}c@{\hspace{.9mm}}
 |@{\hspace{.9mm}}c@{\hspace{.9mm}}	
 |@{\hspace{.9mm}}c@{\hspace{.9mm}}
 |@{\hspace{.9mm}}c@{\hspace{.9mm}}
 |@{\hspace{.9mm}}c@{\hspace{.9mm}}|
}
\hline 
$ q_d $ & $ q_{\varphi} $ & $ q_{\theta} $ & $ q_{\alpha} $ & $ q_{\phi} $ & $ \Delta T $  \\ \hline
 $8.81\,\text{m}^2/\text{s}^4 $  &  $ 3\times 10^{-3} \,\text{rad}^2/\text{s}^4 $  &  $ 1.56\times 10^{-4} \, \text{rad}^2/\text{s}^4 $  & 0 & $ 10^{-6}\,\text{rad}^2/\text{s}^4 $ & $3.3\times10^{-3}\,\text{s}$ \\ \hline
\end{tabular}\\
\vspace{5mm}
\begin{tabular}{
 |@{\hspace{.9mm}}c@{\hspace{.9mm}}
 |@{\hspace{.9mm}}c@{\hspace{.9mm}}
 |@{\hspace{.9mm}}c@{\hspace{.9mm}}
 |@{\hspace{.9mm}}c@{\hspace{.9mm}}|
}
\hline
reinit. of $ \bm{\gamma}_{k,n} $ & estim. of noise/DMC & $ K_{\text{max}} $ & $ \beta_{\text{max}} $ \\ \hline
 $36$th time instance & $5$th time instance & 30  &  40\%  \\ \hline
\end{tabular}
\end{table*}

Clutter components around some high-power MPCs are observed during the tracking. They usually have similar angles and propagation distances as the dominant MPCs close by and experience very short lifetime. These components are mainly generated due to power compensation in the estimation procedure and do not have actual physical meaning, therefore they are not considered in the localization step.

\begin{figure}[t!]
\centering
\subfloat[]{
	\label{subfig:losdistance}
	\includegraphics[width=0.45\textwidth]{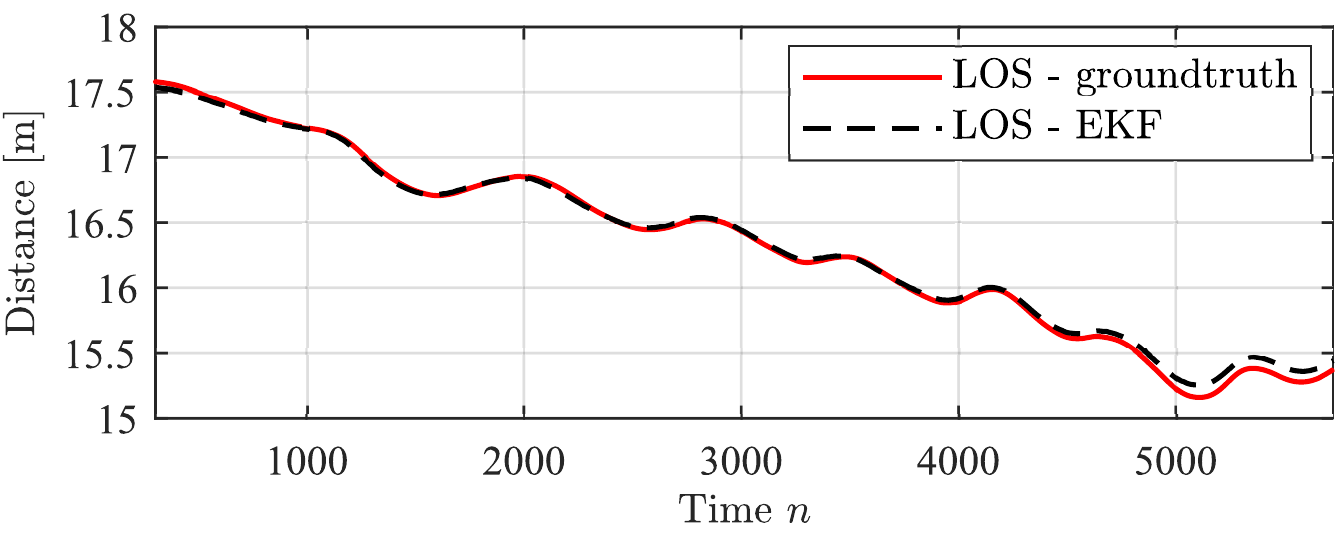}}
	\hspace{2mm}
\subfloat[]{
	\label{subfig:loserror}
	\includegraphics[width=0.45\textwidth]{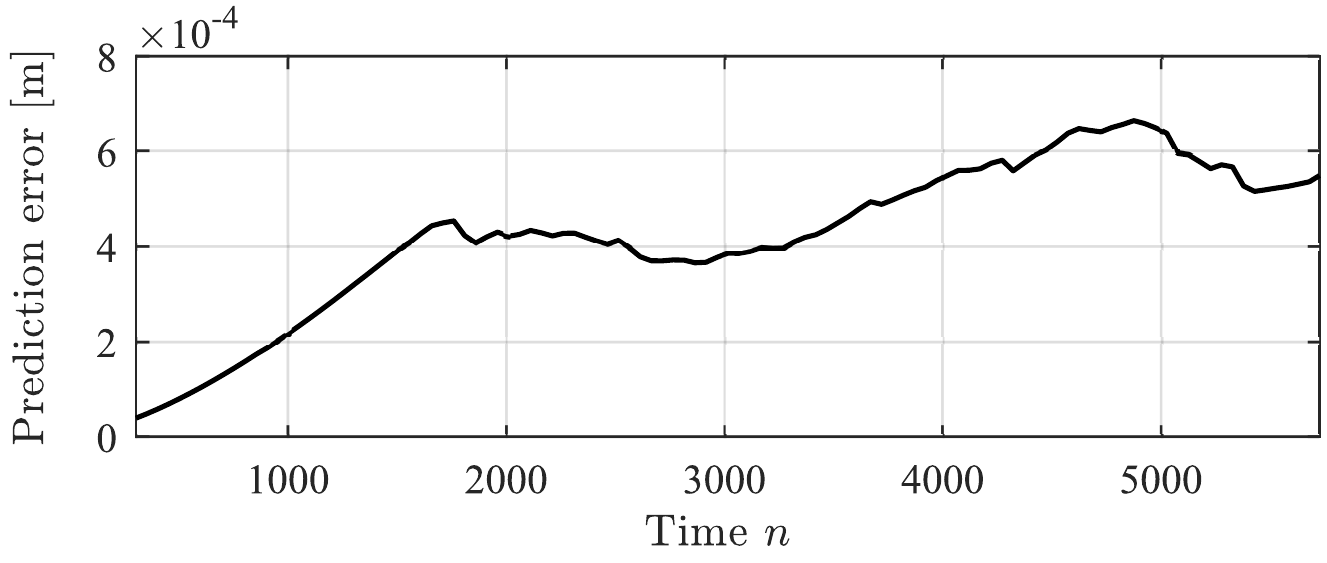}} 
\caption{Performance evaluation of the tracked LOS component. In (a), the black dashed line denotes the distance estimates from EKF. The red solid line is the propagation distance computed with the groundtruth. The two curves are manually synchronized for better comparison. (b) shows the estimation errors of the propagation distances, which are subtracted from the posterior filter covariance matrix.}
\label{kohler} 
\end{figure}

For a better evaluation of the tracking performance, we zoom into the LOS component and compare the distance estimates with the groundtruth. The red solid line in Fig.~\ref{subfig:losdistance} represents the true propagation distance of the LOS path, which is calculated based on the 3D coordinates from the optical system and the groundtruth coordinates of the PA. As shown from the comparison, the EKF performed a smooth tracking of the movements, with all the non-linear and quick motions being captured. The distance estimates have a good match with the groundtruth most of the time, besides some deviations observed after 16\,s. The maximum deviation from the groundtruth is about 8\,cm, while the predicted errors of the LOS distance estimates from the posterior covariance matrix (Fig.~\ref{subfig:loserror}) are in the scale of sub-centimeters, which are clearly underestimated. Also, it is shown that the errors are accumulated during the tracking and reach the maximum at the sharp turns of ``\textbf{L}''. 

\subsubsection{MPC Lifetime Analysis}
In this section, we focus on the statistical characterization of MPC lifetimes in this path intense environment and the analysis is presented from two perspectives: (i) empirical distribution of tracked MPC lifetime and the comparison with statistical distributions, (ii) the relation between the lifetime and parameters like averaged SINR, averaged power of each MPC. The clutter components which do not contain any geometrical meanings are excluded from the statistical analysis. From the phase evolution perspective, we have the minimum resolvability of one wavelength movement, therefore any MPCs with lifetime (converted into distance) less than one wavelength are considered as clutters.

\begin{figure}[t!]
	\centering
	\includegraphics[width=0.45\textwidth]{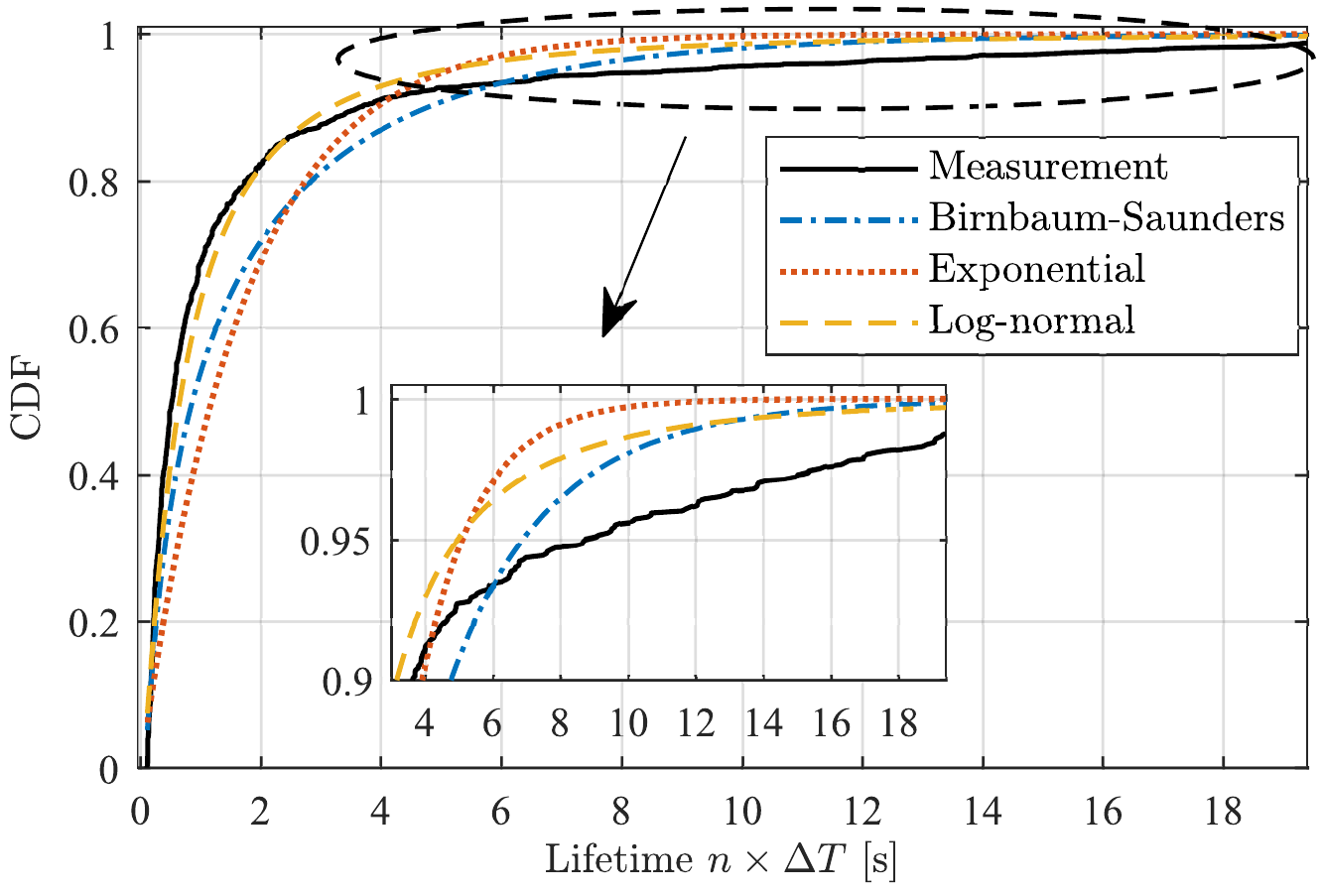}
	\caption{Empirical distribution of tracked MPC lifetime from the ``\textbf{Lund}'' measurement and the comparison with statistical distributions.}
	\label{cdflifetime}
	\vspace{0pt}
\end{figure}

The empirical lifetime cumulative distribution function (CDF) of the tracked MPCs (Fig.~\ref{cdflifetime}) shows that over 90\,$\%$ of the tracked MPCs are with lifetime smaller than 4\,s, and insufficient samples leads to a non-smooth curve from 4\,s to 19.7\,s. The lack of long and robustly tracked MPCs 
clearly make Problem \ref{prob_optmisstoa} in (\ref{eq:optim}) a tougher problem. Further, we considered the lognormal, exponential, and the Birnbaum-Saunders (B-S) \cite{7676244} distributions as the potential fitting statistical distributions for the empirical lifetime CDF, and conducted the goodness-of-fit $ \chi^2 $-test. The lognormal distribution yields a better fit with the empirical CDF especially in the small lifetime region, while the B-S and exponential distributions deviate significantly from the empirical curve. The $ \chi^2 $-tests yield a rejection rate of 100\,$\%$ for all the three distributions. Besides, the mean square error (MSE) of the B-S and exponential distribution compared with the measurement are 0.0192 and 0.0378, respectively, lognormal distribution has the MSE of 0.0038. The significance level is set to 5\,$\%$.

\begin{figure}[t!]
\centering
\subfloat[]{
	\label{PowLifetimeDis}
	\includegraphics[width=0.45\textwidth]{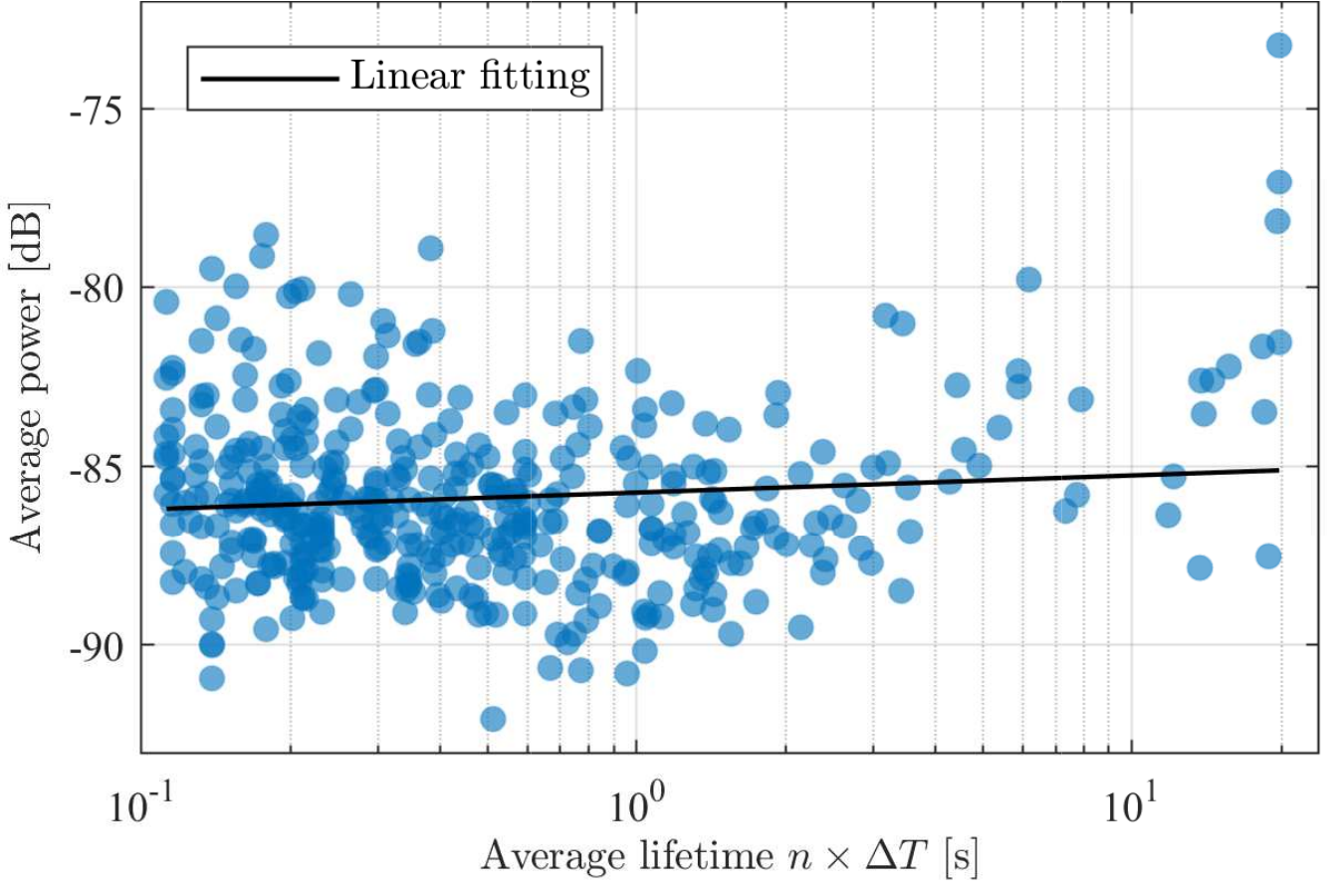}}
	\hspace{2mm}
\subfloat[]{
	\label{SINRLifetimeDis}
	\includegraphics[width=0.45\textwidth]{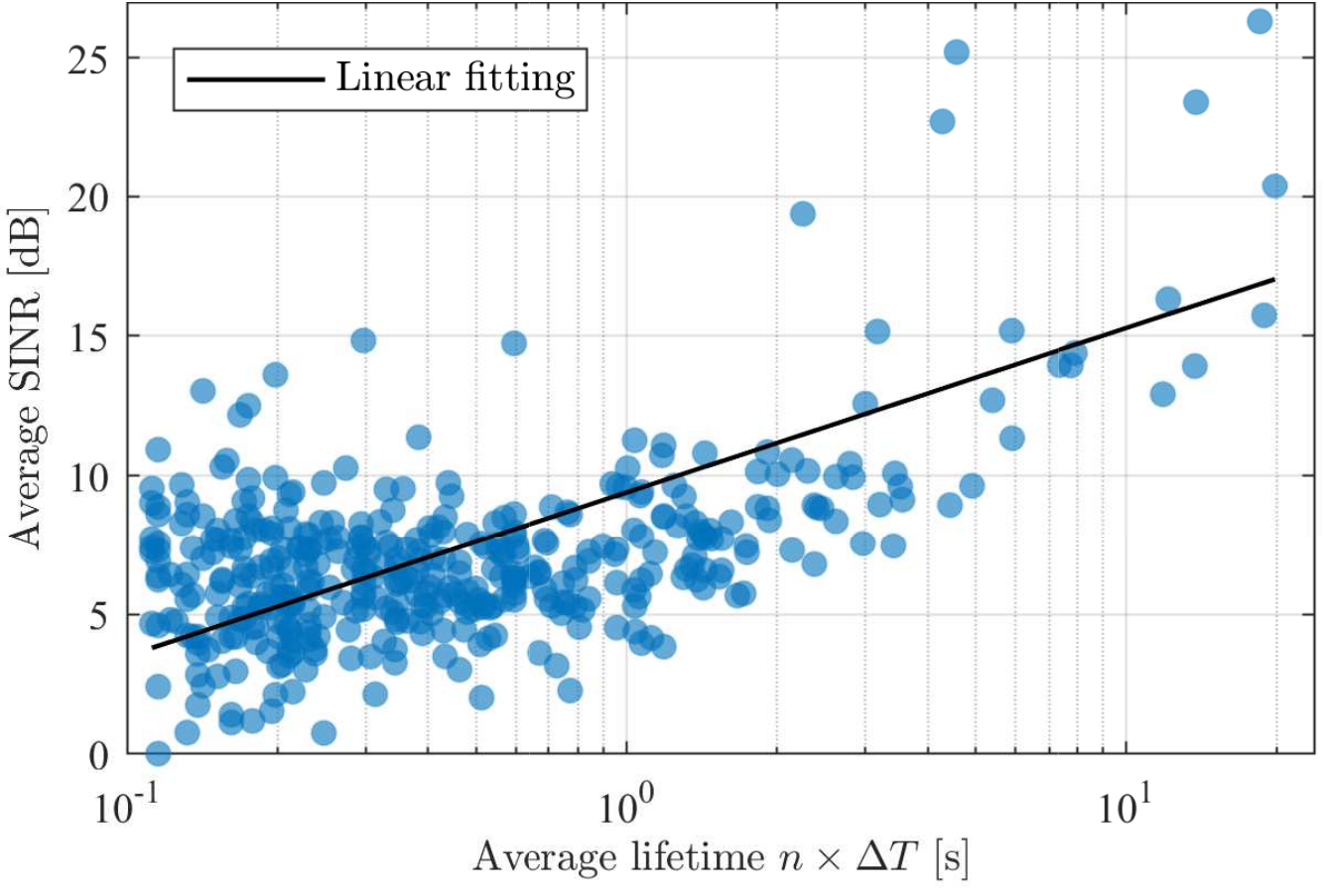}} 
\caption{Relation between the averaged lifetimes [s] and (a) averaged powers [dB] and (b) averaged SINRs [dB] of the tracked MPCs.}
\label{kohler} 
\end{figure}

As shown in (Fig.~\ref{PowLifetimeDis}), the Pearson's rank correlation between the averaged MPC lifetimes and the averaged powers is 0.167, showing a weak dependency between the two variables, which means high power of MPCs does not guarantee continuously stable tracking. The Pearson's rank correlation between the averaged MPC lifetimes and the averaged SINRs is 0.731 (Fig.~\ref{SINRLifetimeDis}), indicating a high dependency between the two variables.

\subsection{Multipath-Assisted Localization}
\label{s5_3}

As seen in Fig.~\ref{MPCdistance} and Fig.~\ref{subfig:losdistance}, most of the specular MPCs can only be observed during fractions of the measurement time (i.e., missing data) and the estimation quality of MPC dispersion parameters is not consistent during the whole tracking process for an individual MPC, i.e., outliers exist in the data, of which the errors are substantial. In this section, we present the performance evaluation of the two experiments described above, with the presence of missing data and outliers.

\subsection{Evaluation of Experiment \RomanNumeralCaps{1}}
We start by looking at experiment \RomanNumeralCaps{1} (\ref{ExpI}), i.e., all the mobile agent positions are assumed to be known, but the inlier set, the feature positions are all unknown. 

\begin{figure}[t!]
	\centering
	\includegraphics[width=0.45\textwidth]{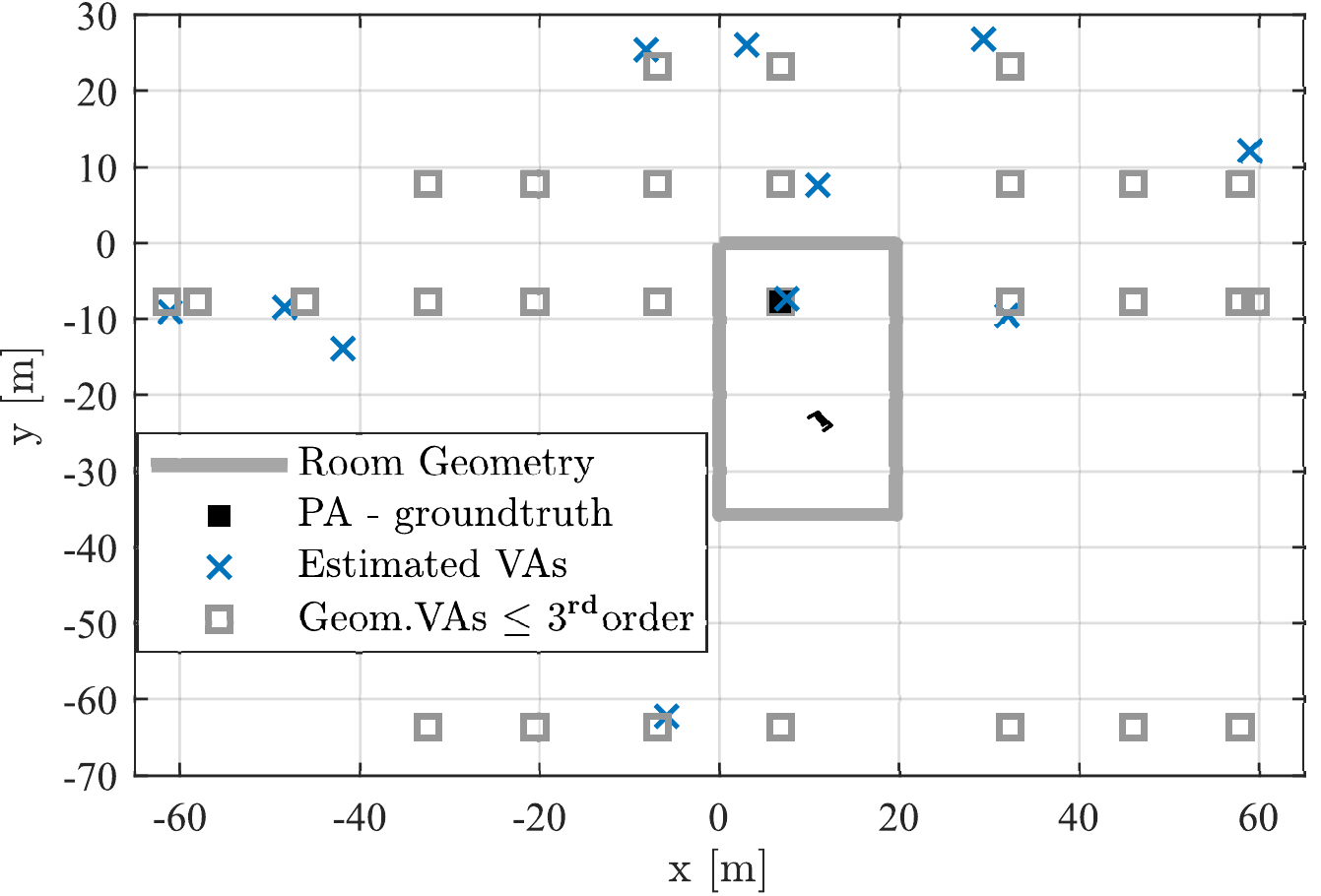}
\caption{Evaluation of experiment \RomanNumeralCaps{1} (\ref{ExpI}). Estimation of feature positions with the prior knowledge of the groundtruth mobile agent positions, meanwhile the inlier set was estimated. Some examples of the estimated and transformed VAs are denoted with blue cross, and the gray squares indicate geometrically expected VAs.}
	\label{fig:planar_lund_rec}
	\vspace{0pt}
\end{figure}
Those tracked MPCs that were longer than $500$ time instances are selected from the tracked 282 MPCs, which gave a set of 50 MPCs. For each of them, we estimated the feature (both PA and VA) positions using RANSAC (to obtain $\hat{I}_{\text{inl}}$) followed by the non-linear optimization of (\ref{eq:optim2}) (to obtain $\hat{\bm{a}}_{k}$). In total these 50 tracked MPCs gave us $103\, 480$ distance samples, i.e., approximately $2000$ each. Of these $77\, 490$ were considered to be inliers. This gives us an estimated inlier ratio of $75\%$. The standard deviation of the inlier residuals is 4.6\,cm. Some examples of the estimated VA positions which corresponds to some long tracked MPCs are shown in Fig.~\ref{fig:planar_lund_rec}, where the reconstructed PA position $\hat{\bm{a}}_{1}$ is registered to the groundtruth PA position, meanwhile the same transformation is applied to all the estimated VA positions. It could be observed that the estimated and transformed VA positions reasonably reconstructed the geometrical features, even for the $ 2^{\text{nd}} $ and $ 3^{\text{rd}} $ order VAs.  



\subsection{Evaluation of Experiment \RomanNumeralCaps{2}}
\begin{figure}[t!]
	\centering
	\includegraphics[width=0.45\textwidth]{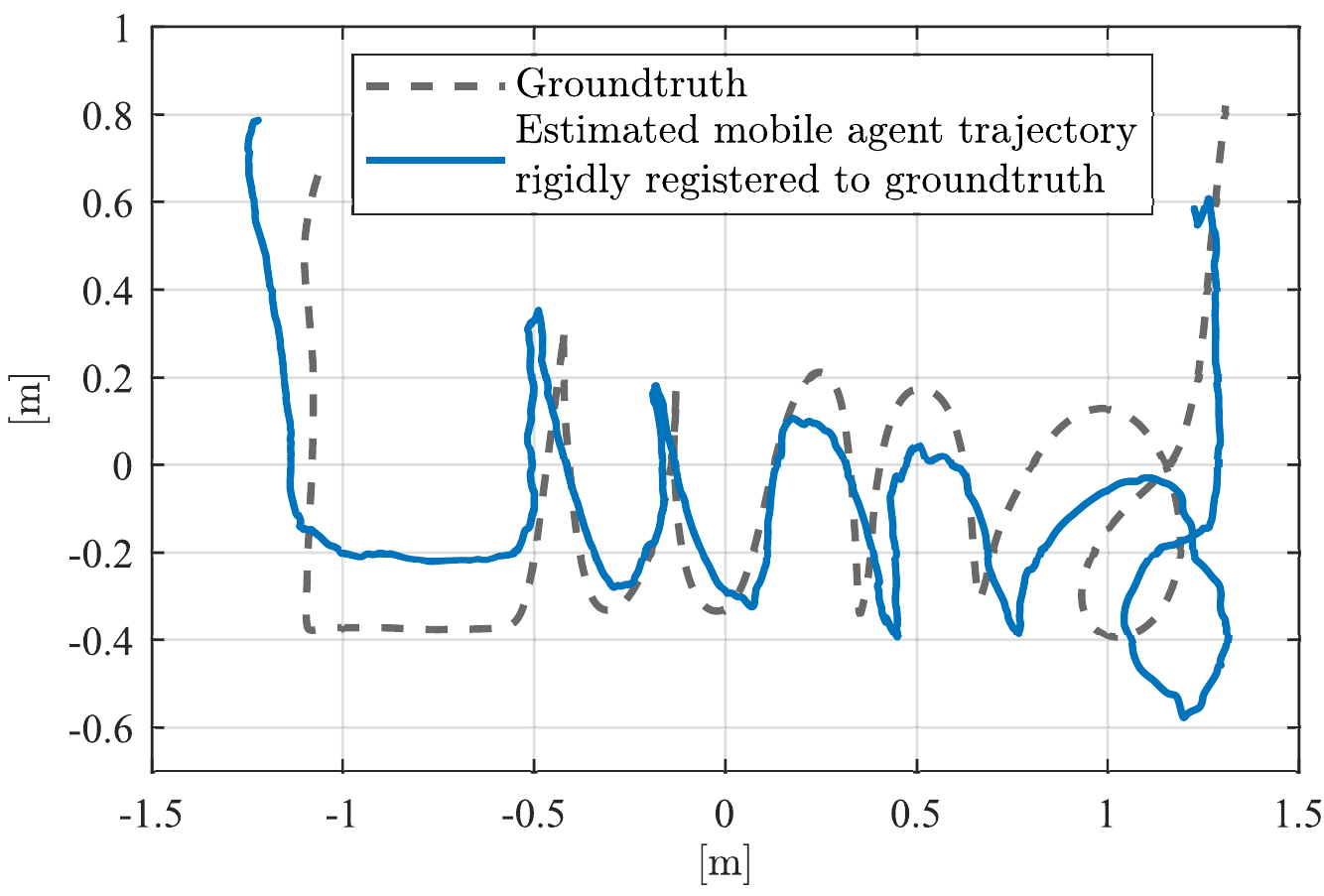}
	\caption{Evaluation of experiment \RomanNumeralCaps{2} (\ref{ExpII}). The groundtruth (deshed gray), and the estimated mobile agent trajectory (solid blue) which is registered to the groundtruth coordinate system.}
	\label{fig:ground_truth_vs_estimated_tikz}
	\vspace{0pt}
\end{figure}

We now turn our attention to the experiment \RomanNumeralCaps{2} (\ref{ExpII}), where only the distance estimates $ \hat{d}_{k,n} $ are given as input and no prior knowledge about the mobile agent positions and feature positions. In order to use the calibration procedure described in the previous section, for the ``\textbf{Lund}'' dataset, we proceeded by splitting the whole dataset in a number of smaller segments in time. This resulted in 117 segments of length 100 time instances with 50 time-instances overlap between adjacent segments. For each segment, we initialized both $ \hat{\bm{a}}_{k} $ and $ \hat{\bm{p}}_n $ using the RANSAC in Section~\ref{ExpII}. The different solutions from the 117 segments were then registered into a common coordinate system using the overlap between the segments. The estimated mobile agent trajectory and the groundtruth are shown in Fig.~\ref{fig:ground_truth_vs_estimated_tikz}.

Considering the estimated mobile agent trajectory and the groundtruth are in different coordinate systems, the alignment between the two systems is firstly needed for further performance evaluation. This is done by solving the following least-squares problem
\begin{align}
\min_{\bm{R},\bm{r}_{0}} \sum_i || \bm{R}\hat{\bm{p}}_i+\bm{r}_{0}-\bm{p}_{\text{true},i} ||^2,
\label{eq:alignTrajectories}
\end{align}
where $ \bm{R} $ is the rotation matrix, and $ \bm{r}_{0} $ is the translational offset vector \cite{horn1988closed}, \cite{kabsch1978discussion}. It could be observed that the estimated trajectory shows a clear ``\textbf{Lund}'' pattern, with all the fine movements details caught. However, the overall shape is stretched along the diagonal direction, which results in a larger deviation from the groundtruth especially in the beginning and the end. The largest deviation of the estimated mobile agent position from the groundtruth happens at the sharp turn of ``\textbf{L}'', which is 26\,cm. Furthermore, the root mean square error (RMSE) is defined as $ d_{\text{RMSE}} = \sqrt{(\sum_{n=1}^{N}|\hat{\bm{p}}_{n}-\bm{p}_{\text{true},n}|^2)/N} $, and the RMSE of the estimated agent trajectory (after being registered) compared with the groundtruth is 14\,cm.



\section{Conclusion and Outlook}
\label{s6}
In this paper, we introduced a high-resolution phase-based localization and mapping framework using massive MIMO system. The proposed channel estimation and tracking algorithm uses an EKF and tightly couples the phase-based distance to the phase shift between consecutive channel measurements, which makes it possible to resolve the MPC distances accurately even when using only low signal bandwidth. A distance-based localization and mapping algorithm is then used for the mobile agent trajectory estimation with the presence of missing data and outliers. The performance evaluation with a real indoor measurement shows that the proposed localization framework can achieve outstanding accuracy even with standard cellular bandwidths. The largest agent position error is 26\,cm and the RMSE position error is 14\,cm. Besides, no prior knowledge of the surroundings and base station position is needed, hence the framework can be applied in different environments given that there are sufficiently many scatterers present. To sum up, the multipath-distance-based localization method that exploits the phases of MPCs using massive MIMO is a promising high-resolution localization solution for current and next generation cellular systems. 

Regarding the future research, the current localization algorithm can be extended to further exploit MPC parameters like AoAs/AoDs, while the array orientation information is needed to calibrate angular estimates into the global coordinate system. Moreover, a soft-decision association between the estimated MPCs and environment features using probabilistic approach can be used to replace the hard-decision association now.




\section{Acknowledgements}
\label{s7}
The authors would like to thank Jose Flordelis, Joao Vieira and Christian Nelson for helping with the measurements, as well as the support from Bj\"{o}rn Olofsson with the motion capture system. We would also like to acknowledge the fruitful discussions within framework of the COST Action CA15104 (IRACON).

\bibliographystyle{IEEEtran}
\bibliography{ref}

\end{document}